\documentclass[journal]{IEEEtran}
\usepackage{amsmath,amsfonts}
\usepackage{algorithmic}
\usepackage{epsfig, algorithm}
\usepackage{array}
\usepackage{amssymb}
\usepackage{acronym} 
\usepackage[caption=false,font=normalsize,labelfont=sf,textfont=sf]{subfig}
\usepackage{textcomp}
\usepackage{stfloats}
\usepackage{tabularx}
\usepackage{xcolor}
\usepackage{url}
\usepackage{verbatim}
\usepackage{multirow}
\usepackage{acronym}
\usepackage{graphicx}	
\usepackage{stfloats}
\usepackage{soul}
\usepackage{cancel}
\def\BibTeX{{\rm B\kern-.05em{\sc i\kern-.025em b}\kern-.08em
T\kern-.1667em\lower.7ex\hbox{E}\kern-.125emX}}
\usepackage{balance}
\usepackage{multirow}
\usepackage{hhline}
\input{AcronymsListFinal}
\acresetall

\usepackage{fancyhdr}

\begin{document}
\title{Optimal Joint Radar and Communication User Association in Cell-Free mMIMO Systems}
%%%%%%%%%%%%%%%%%%%%%%%%%%%%%%%%%%%%%%%%%%%%%%%%%%%%%%%%%%%%%%%%%%%%%%%%%%%%%%%%%
\author{Ahmed Naeem, El Mehdi Amhoud~\IEEEmembership{Member,~IEEE}, Hüseyin Arslan~\IEEEmembership{Fellow,~IEEE}% <-this % stops a space
\thanks{Ahmed Naeem and Hüseyin Arslan are with the Department of Electrical and Electronics
Engineering, Istanbul Medipol University, Istanbul, 34810, Turkey (email: ahmed.naeem@std.medipol.edu.tr, huseyinarslan@medipol.edu.tr).}% <-this % stops a space
\thanks{El Mehdi Amhoud is with the College of Computing, Mohammed VI Polytechnic University, Benguerir, Morocco (email: elmehdi.amhoud@um6p.ma).}
\\This work has been submitted to the IEEE for possible publication. Copyright may be transferred without notice, after which this version may no longer be accessible.}
%%%%%%%%%%%%%%%%%%%%%%%%%%%%%%%%%%%%%%%%%%%%%%%%%%%%%%%%%%%%%%%%%%%%%%%%%%%%%%%%%
\maketitle 
%%%%%%%%%%%%%%%%%%%%%%%%%%%%%%%%%%%%%%%%%%%%%%%%%%%%%%%%%%%%%%%%%%%%%%%%%%%%%%%%%
\begin{abstract}
The cell-free massive multiple-input multiple-output (CF-mMIMO) systems are crucial for 6G development due to their high spectral efficiency and uniform user-experienced data rates. A key aspect of CF-mMIMO is user association (UA) and optimal cluster formation. Traditional methods focusing solely on communication-related metrics fall short in this context, as sensing is becoming integral to 6G. This study delves into a framework for joint radar and communication (JRC) in CF-mMIMO systems and investigates JRC-based UA techniques. We propose a novel method to optimize UA, enhancing both communication spectral efficiency and sensing accuracy. Existing literature has not explored this dual requirement integration for UA. Our proposed two-step scheme optimizes UA clusters for both communication and sensing. The first step involves selecting access points (APs) based on channel quality, followed by a second step that further refines the selection by choosing APs from the initial group that are also optimal for sensing. We utilize the signal-clutter plus noise ratio to exclude APs with clutter in front of the user equipment (UE) and the AP view angle, ensuring that radar echoes are received only from the specific UE, not the surrounding clutter. Theoretical analysis and simulations demonstrate that the same APs optimized for communication are not necessarily optimal for sensing, highlighting the need for schemes that incorporate sensing requirements in UA. The results show the effectiveness of the proposed method, showing its potential to improve CF-mMIMO system performance in JRC scenarios.
\end{abstract}

%%%%%%%%%%%%%%%%%%%%%%%%%%%%%%%%%%%%%%%%%%%%%%%%%%%%%%%%%%%%%%%%%%%%%%%%%%%%%%%%%
\begin{IEEEkeywords}
Cell-Free mMIMO, clutter, joint radar and communication, sensing, spectral efficiency, user association.
\end{IEEEkeywords}
%%%%%%%%%%%%%%%%%%%%%%%%%%%%%%%%%%%%%%%%%%%%%%%%%%%%%%%%%%%%%%%%%%%%%%%%%%%%%%%%%
\IEEEpeerreviewmaketitle
\section{INTRODUCTION}
\par With the ever-growing connectivity, \ac{6G} will revolutionize networks by offering high data rates, ultra-low latency, and seamless connectivity through advancements like higher frequencies, \ac{mMIMO}, and intelligent network management. Furthermore, in conjunction with wireless communication developments, sensing will also play a pivotal role in enhancing \ac{6G} leading to the emergence of \ac{JRC}, with many applications such as virtual reality, autonomous vehicles, and activity recognition \cite{wei2023integrated}. Thus, integrating these technologies is crucial for realizing the full potential of \ac{6G}.
\par Building upon the above advancements, improving service quality and uniform data rates in future networks is pivotal. Current networks achieve high peak data rates at cell centers but face substantial variations and inconsistent service quality at cell edges \cite{andrews2016we}. Despite deploying ultra-dense networks and \ac{mMIMO} systems to enhance user-experienced data rates, challenges like significant \ac{SNR} fluctuations and inter-cell interference persist \cite{8258595}. Future networks should ensure consistent user-experienced data rates across the coverage area, not just increasing peak or average rates \cite{chen2022survey}.
\par The \ac{CF-mMIMO} is a promising approach to address the challenges of inconsistent service quality and data rates faced by current networks. This technology combines aspects of both ultra-dense networks and \ac{mMIMO} systems to ensure uniform data rates across the coverage area \cite{demir2021foundations}. The main motivation is to enhance user-experienced data rates, rather than merely increasing peak rates \cite{ngo2015cell}. A \ac{CF-mMIMO} can be visualized as a network with a dismantled \ac{mMIMO} array \cite{10411070}, where individual antennas are deployed at various locations. For a specific \ac{UE}, the distributed antennas transmit data signals with varying power and phase shifts, ensuring that the signals reach the \ac{UE} synchronously and reinforce each other \cite{chen2022survey}.
\par One of the critical factors in the effectiveness of \ac{CF-mMIMO} is \ac{UA} \cite{bjornson2020scalable}. Unlike the conventional cellular networks where each \ac{UE} is only associated with a single \ac{AP}, in \ac{CF-mMIMO} each \ac{UE} is served by a subset of \ac{APs}. Consequently, research in the field of \ac{UA} in \ac{CF-mMIMO} has gained significant attention for its potential to optimize \ac{UE} and \ac{AP} associations, thereby enhancing reliability, data rates, and overall system performance. When considering \ac{AP} selection schemes in \ac{CF-mMIMO} systems for communication-only fall into two categories: large-scale-based and competition-based schemes. In large-scale-based schemes, \ac{UEs} select \ac{APs} with the largest \ac{LSFC} and better channel conditions. Each \ac{LSFC} is sorted in order and compared to a threshold \cite{chen2022survey}. A user-centric virtual cell approach is introduced in \cite{8000355,chen2020structured}, where \ac{UA} is done by considering the strongest channels (i.e., largest norm). Moreover, in \cite{ngo2018performance} an \ac{AP} selection scheme is introduced subject to reducing the backhaul requirements. A joint \ac{AP} and \ac{UE} preference-based scheme to ensure scalability for all \ac{UEs} is proposed in \cite{sarker2023access}. Another association method is proposed in \cite{d2020user}, which uses the Hungarian algorithm to create a cluster of \ac{APs} to serve a \ac{UE} based on their locations. Furthermore, in a competition-based scheme, a new accessing \ac{UE} competes for an \ac{AP} already serving its maximum number of \ac{UEs}. The \ac{AP} prioritizes \ac{UEs} with the best channel conditions by identifying the weakest \ac{UE} it serves. If the new \ac{UE} offers a better channel, it replaces the weakest one, which then blacklists the \ac{AP} \cite{chen2022survey,chen2020structured}. 
\par All the aforementioned works primarily focus on associating \ac{UEs} with \ac{APs} from the perspective of communication-related service metrics only. However, to meet the dual needs of communication and sensing, new optimized \ac{UA} schemes are required to enhance overall \ac{JRC} network performance. For an optimized \ac{UA} scheme in a \ac{JRC} \ac{CF-mMIMO} system, robust communication links are essential to achieve high data rates and reliable connectivity. Simultaneously, accurate sensing is vital for applications such as localization, target tracking, environmental monitoring, and autonomous systems. Therefore, integrating both aspects into \ac{UA} strategies is crucial. Moreover, the \ac{CF-mMIMO} architecture is particularly advantageous for multi-static sensing, providing high resolution, robust sensing, and a wider range of sensing angles \cite{behdad2024multi}. Research on \ac{JRC} in \ac{CF-mMIMO} systems is limited, existing work mainly focuses on power allocation \cite{behdad2024multi,behdad2024joint}, multi-target detection \cite{elfiatoure2024multiple,behdad2024multi}, \ac{JRC} beamforming \cite{mao2023beamforming}, and defining communication and sensing regions \cite{mao2024communication}. Despite these research efforts, the field of \ac{JRC} in \ac{CF-mMIMO} systems remains an active area of investigation, with significant potential for further advancements.
\par To effectively address communication and sensing needs, this paper highlights a new approach for optimizing \ac{JRC}-based \ac{UA} in \ac{CF-mMIMO}. The key contributions are
\begin{enumerate}
\item This paper proposes a \ac{JRC} framework for \ac{CF-mMIMO} systems, where each \ac{AP} performs sensing and communication concurrently. The proposed framework provides a detailed step-by-step explanation of the signaling involved, offering a comprehensive understanding of the \ac{JRC} \ac{CF-mMIMO} architecture.
\item Unlike previous studies that have not integrated sensing requirements into \ac{CF-mMIMO} \ac{UA}, this paper proposes a novel two-step approach that combines both communication and sensing requirements. By jointly addressing these requirements, the proposed \ac{UA} scheme aims to enhance the overall performance of the \ac{JRC} \ac{CF-mMIMO} systems.
\item This dual-step approach leverages \ac{SCNR} to filter out \ac{APs} from the cluster that are affected by significant clutter, ensuring precise radar echoes only from specific \ac{UE}(s). The proposed scheme first selects \ac{APs} based on channel quality and then refines this selection by considering \ac{APs} that are also optimal for sensing. The paper also provides detailed optimization problems and corresponding algorithmic solutions to address this issue.
\item We analyze the communication \ac{SE}, the impact of clutter on \ac{$P_{dc}$}, and how different clutter densities and \ac{RCS} of the clutter affect the performance of the \ac{UA}. A detailed mathematical analysis is provided in both \ac{LoS} and \ac{NLoS} radar echo reception in a cluttered environment and its impact on the performance of sensing and \ac{AP} selection. 
\item Moreover, we conduct detailed simulations and theoretical analysis, including detailed mathematical evaluations, to assess the performance of the proposed \ac{UA} scheme. The results show significant improvements in communication and sensing compared to traditional methods, validating the effectiveness of our approach. Our findings highlight the importance of accounting for sensing requirements in a cluttered environment to enable optimal \ac{UA} decisions in the integrated \ac{JRC} \ac{CF-mMIMO} systems.
\end{enumerate}
The paper is organized as: Section II presents the system model and framework of \ac{JRC} \ac{CF-mMIMO}. Section III formulates the optimization problems. The proposed \ac{UA} method is detailed in Section IV, followed by the simulation results in Section V. Finally, Section VI concludes the paper. \footnote{\textit{Notation:} Boldface uppercase letters, $\mathbf{X}$, denote matrices, and boldface lowercase letters, $\mathbf{x}$, denote column vectors, Superscripts $^\mathrm{T}$, $^{*}$, and $^\mathrm{H}$ denote transpose, conjugate, and conjugate transpose, respectively.
The entry $(i,j)$ of matrix $\mathbf{X}$ is $\mathbf{X}_{ij}$, and $\mathbf{X}_{\cdot j}$ is its $j$-$th$ column. The $n\times n$ identity matrix is $\mathbf{I}_n$. We use $\triangleq$ for definitions and $\operatorname{diag}\left(\mathbf{A}_1,\ldots,\mathbf{A}_n\right)$ for a block-diagonal matrix with square matrices $\mathbf{A}_1,\ldots,\mathbf{A}_n$ on the diagonal. The multivariate circularly symmetric complex Gaussian distribution with correlation matrix $\mathbf{R}$ is denoted $\mathcal{N}_\mathbb{C}\left(\mathbf{0},\mathbf{R}\right)$. The Euclidean norm of $\mathbf{x}$ is $\left\|\mathbf{x}\right\|_2$. The expected value of $\mathbf{x}$ is $\mathbb{E}\left\{\mathbf{x}\right\}$. We denote the cardinality of set $\mathcal{A}$ by $|\mathcal{A}|$ and its $n$-$th$ element by $\mathcal{A}(n)$.}
%%%%%%%%%%%%%%%%%%%%%%%%%%%%%%%%%%%%%%%%%%%%%%%%%%%%%%%%%%%%%%%%%%%%%%%%%%%%%%%%%
\section{System Model} \label{sysModel}
\par This section explains the \ac{JRC}-based \ac{UA} framework in \ac{CF-mMIMO}, with a \ac{DFRC} \cite{9940605}, monostatic topology where a group of \ac{APs} communicates with a \ac{UE} while simultaneously sensing and tracking its movement \cite{9724187}. Unlike \cite{behdad2024multi}, where each \ac{AP} is either a \ac{JRC} transmitter or a sensing receiver, our approach enables each \ac{AP} to function as both, for better resources utilization.
%%%%%%%%%%%%%%%%%%%%%%%%%%%%%%%%%%%%%%%%%%%%%%%%%%%%%%%%%%%%%%%%%%%%%%%%%%%%%%%%% 
\begin{figure}
\centering 
\resizebox{0.8\columnwidth}{!}{
\includegraphics{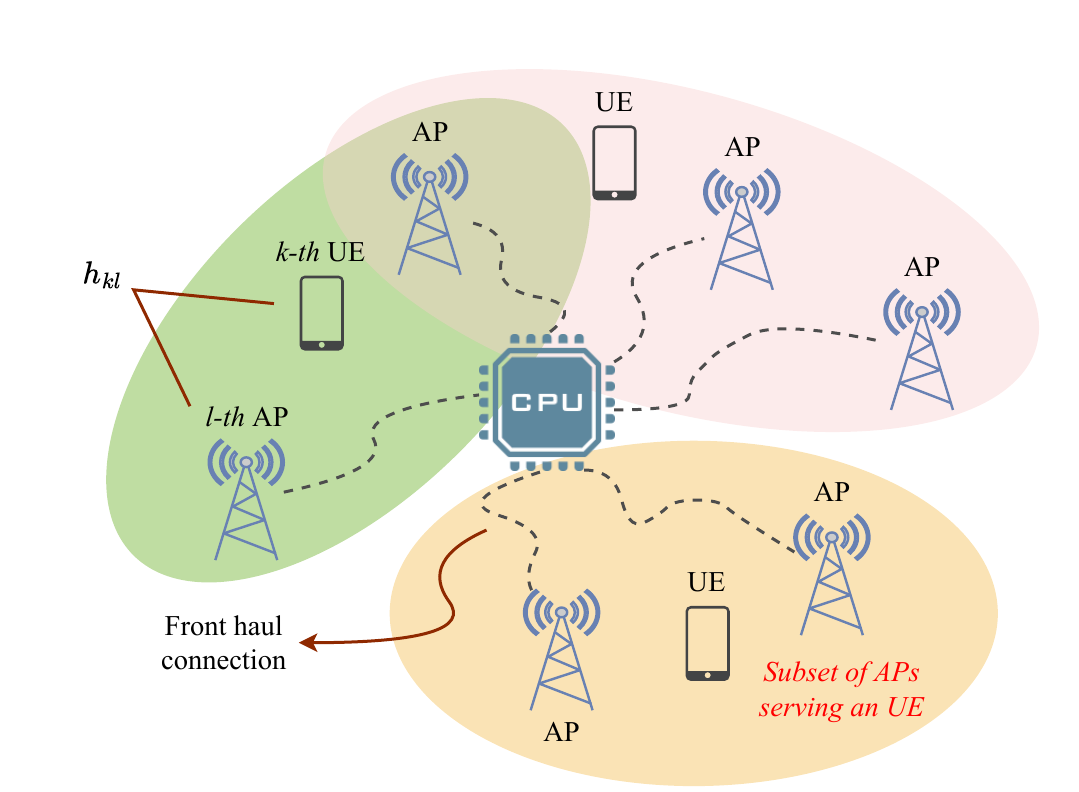}}
\caption{Cell-Free mMIMO network.}
\label{cellfree}
\end{figure}
%%%%%%%%%%%%%%%%%%%%%%%%%%%%%%%%%%%%%%%%%%%%%%%%%%%%%%%%%%%%%%%%%%%%%%%%%%%%%%%%%
\vspace{-2mm}
\subsection{Network Architecture}
\par
\begin{figure}
\centering 
\resizebox{1\columnwidth}{!}{
\includegraphics{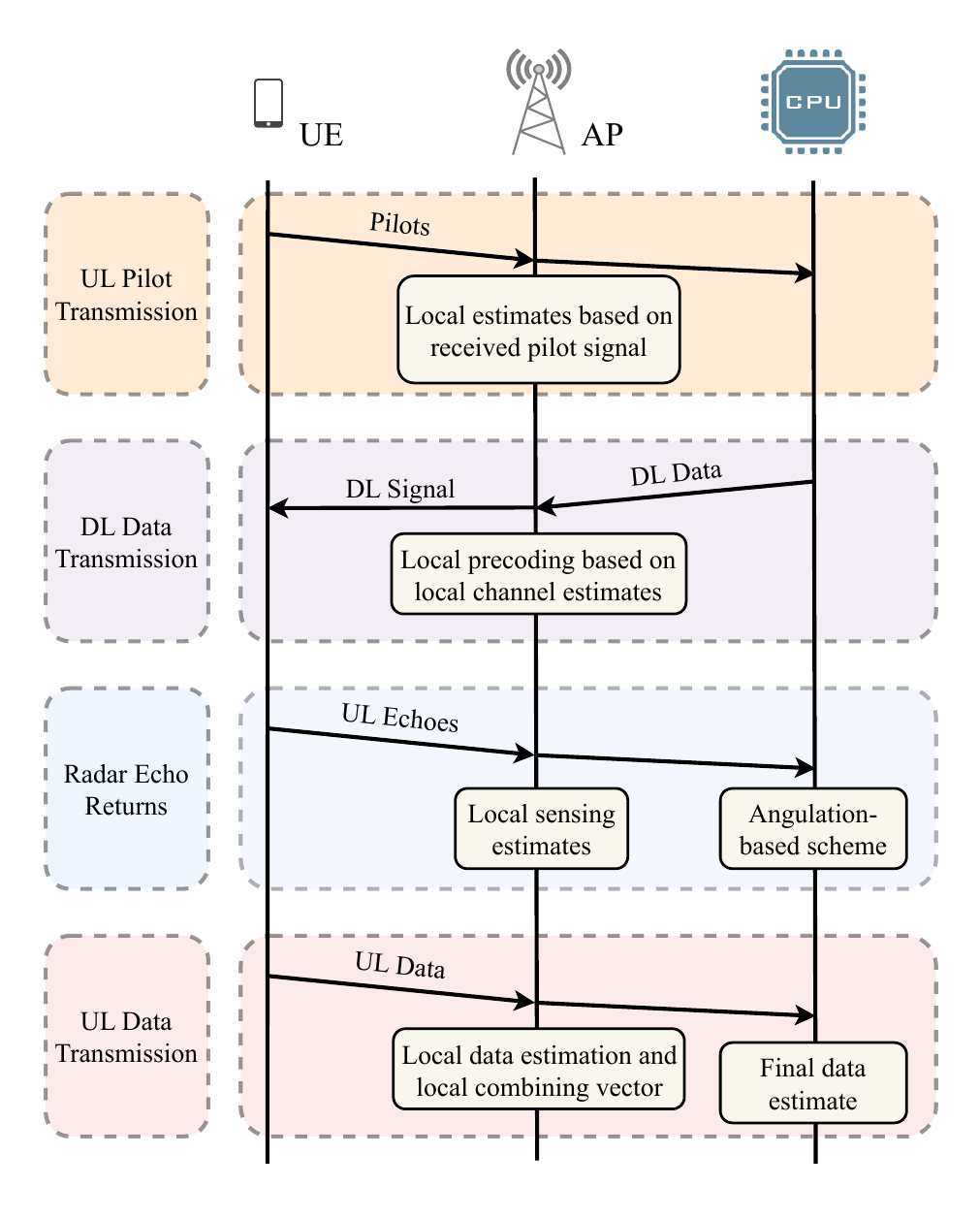}}
\caption{Proposed distributed signaling scheme.}
\label{Signal}
\end{figure}

 We consider a \ac{CF-mMIMO} with $K$ single-antenna \ac{UEs}, $\mathcal{K} \triangleq \{1,2,\ldots,K\}$, and $L$ randomly located \ac{APs} with $N$ antennas for 3D beamforming as \cite{10058895} and \cite{9713691}. All \ac{APs} are connected via a front-haul to a \ac{CPU}, as shown in Fig. \ref{cellfree}. We assume a distributed implementation, following \ac{TDD} protocols as illustrated in Fig. \ref{Signal}. The channel between the $l$-$th$ \ac{AP} and the $k$-$th$ \ac{UE} is $\mathbf{h}_{kl} \in \mathbb{C}^{N \times 1}$ with collective channel from all \ac{APs} as $\mathbf{h}_k=[\mathbf{h}_{k1}^\mathrm{T}\ldots\mathbf{h}_{kL}^\mathrm{T}]^\mathrm{T}\in\mathbb{C}^M$ with $M=NL$, (representing the total number of antennas in the
coverage area). During the coherence block $\tau_c$, $\mathbf{h}_{kl}$ remains constant and follows a correlated Rayleigh fading distribution, $\mathbf{h}_{k l} \sim \mathcal{N}_{\mathbb{C}}\left(\mathbf{0}, \mathbf{R}_{k l}\right)$ \cite{bjornson2017massive}, where $\mathbf{R}_{kl}=\mathbb{E}\left[\mathbf{h}_{kl} \mathbf{h}_{kl}^{\mathrm{H}}\right] \in \mathbb{C}^{N \times N}$ is the spatial correlation matrix, incorporating path-loss, shadowing, and spatial correlation \cite{sarker2023access}. Moreover, the channels from all \ac{APs} are independently distributed, i.e., \(\mathbb{E}\{\mathbf{h}_{kn}(\mathbf{h}_{kl})^{\mathrm{H}}\} = \mathbf{0}\) for \(l \neq n\). The collective distribution is $\mathbf{h}_k\sim\mathcal{N}_\mathbb{C}(\mathbf{0},\mathbf{R}_k),$ where $\mathbf{R}_k=\text{diag}(\mathbf{R}_{k1},\ldots,\mathbf{R}_{kL})\in \mathbb{C}^{M \times M}$ is block-diagonal spatial correlation matrix. The \ac{LSFC} between the $l$-$th$ \ac{AP} and the $k$-$th$ \ac{UE} is denoted by $\beta_{kl}$, which is available at all \ac{APs} and \ac{UEs}. This \ac{LSFC} is the normalized trace of $\mathbf{R}_{kl}$, given by $\beta_{kl}=\operatorname{tr}\left(\mathbf{R}_{kl}\right) / N$, and modeled as \cite{sarker2023access}
%%%%%%%%%%%%%%%%%%%%%%%%%%%%%%%%%%%%%%%%%%%%%%%%%%%%%%%%%%%%%%%%%%%%%%%%%%%%%%%%%
%%%%%%%%%%%%%%%%%%%%%%%%%%%%%%%%%%%%%%%%%%%%%%%%%%%%%%%%%%%%%%%%%%%%%%%%%%%%%%%%%
%%%%%%%%%%%%%%%%%%%%%%%%%%%%%%%%%%%%%%%%%%%%%%%%%%%%%%%%%%%%%%%%%%%%%%%%%%%%%%%%%
\begin{equation} \label{LSFC}
\beta_{kl}(\mathrm{dB}) = \Upsilon - 10 \alpha \log_{10}\left(\frac{d_{kl}}{d_{\text{ref}}}\right) + Z_{kl},
\end{equation}
%%%%%%%%%%%%%%%%%%%%%%%%%%%%%%%%%%%%%%%%%%%%%%%%%%%%%%%%%%%%%%%%%%%%%%%%%%%%%%%%%
where $\Upsilon$ is the path loss at reference distance $d_{\text{ref}}$, $d_{kl}$ is the distance between $l$-$th$ \ac{AP} and $k$-$th$ \ac{UE}, $\alpha$ is the path loss exponent, and $Z_{kl}$ is the shadowing effect, normally distributed with zero mean and standard deviation $\sigma_{dB}$. Finally, for different links of the channel vectors separate and independent realizations are obtained.
\par The original design of \ac{CF-mMIMO} systems, where all \ac{UEs} are served by all \ac{APs} is impractical \cite{ngo2017cell}. To ensure scalability in this case, we incorporate a set of block-diagonal matrices $\mathbf{F}_{kl}=\operatorname{diag}\left(\mathbf{F}_{k1}, \ldots, \mathbf{F}_{kL}\right), k=1, \ldots, K$ and $l=1, \ldots, L$, where the antenna configuration at the $l$-$th$ \ac{AP} for the $k$-$th$ \ac{UE} is a diagonal matrix represented by $\mathbf{F}_{kl} \in \mathbb{C}^{N \times N}$. The $n$-$th$ diagonal entry of $\mathbf{F}_{kl}$ is 1 if the $n$-$th$ antenna of the $l$-$th$ \ac{AP} is used by the $k$-$th$ \ac{UE} and 0 otherwise. Moreover, $\mathbf{F}_{kl}$ determines the matrix $\mathbf{S} \in \mathbb{R}^{K \times L}$, which is defined as the association matrix. This binary matrix $\mathbf{S}$ specifies the \ac{AP} selection between \ac{UEs} and \ac{APs}, where for the $l$-$th$ \ac{AP} and the $k$-$th$ \ac{UE} it is defined as
%%%%%%%%%%%%%%%%%%%%%%%%%%%%%%%%%%%%%%%%%%%%%%%%%%%%%%%%%%%%%%%%%%%%%%%%%%%%%%%%%
\begin{equation}
\mathbf{S}_{kl} = 
\begin{cases} 
1, & \text{if} \ \operatorname{tr}\left(\mathbf{F}_{kl}\right) > 0 \\ 
0, & \text{otherwise}.
\end{cases} 
\end{equation}
%%%%%%%%%%%%%%%%%%%%%%%%%%%%%%%%%%%%%%%%%%%%%%%%%%%%%%%%%%%%%%%%%%%%%%%%%%%%%%%%%
where the entry $\mathbf{S}_{kl} = 1$ if the trace of $\mathbf{F}_{kl}$ is greater than 0, indicating that the $l$-$th$ \ac{AP} serves the $k$-$th$ \ac{UE}. To improve the clarity of the mathematical descriptions, we use the notation $\mathcal{M}_k=\left\{l: \mathbf{S}_{kl}=1, l \in \{1, \ldots, L\}\right\}$ be the subset of \ac{APs} serving the $k$-$th$ \ac{UE}, and $\mathcal{D}_l=\left\{k: \mathbf{S}_{kl}=1, k \in \{1, \ldots, K\}\right\}$ the subset of \ac{UEs} served by the $l$-$th$ \ac{AP}. 
\par The total coherence interval $\tau_c$ is divided into three sub-intervals: (a) \ac{UL} pilots transmission $\tau_p$, to estimate channel between the $k$-$th$ \ac{UE} and the $l$-$th$ \ac{AP}, (b) $\tau_d$ symbols for \ac{DL} \ac{DFRC} signal, through channel-matched beamforming. During this \ac{DL} transmission, radar echoes are received at the \ac{APs} after reflecting back from the \ac{UE}, i.e., $\tau_d = \tau_{d(\mathrm{c})} + \tau_{u(\mathrm{e})}$, where $\tau_{d(\mathrm{c})}$ is the \ac{DL} communication signal and $\tau_{u(\mathrm{e})}$ is the radar echo received at each \ac{AP} and (c) \ac{UL} data symbols $\tau_u$.
%%%%%%%%%%%%%%%%%%%%%%%%%%%%%%%%%%%%%%%%%%%%%%%%%%%%%%%%%%%%%%%%%%%%%%%%%%%%%%%%%
\vspace{-3mm}
\subsection{Uplink Pilot Transmission and Channel Estimation} \label{ULSec}
\par During initial access, \ac{UEs} are randomly assigned pilots from a set of orthogonal signals, $\{\phi_1,\ldots,\phi_{\tau_p}\}$ each with length $\tau_p$ samples, $\phi \in \mathbb{C}^{1 \times \tau_p}$. These pilot signals ensure equal power, satisfying $\left|\phi_t\right|^2=\tau_p$, where $\tau_p$ remains constant and is independent of $K$. A massive access scenario is considered ($K\gg\tau_p$), leading to \textit{pilot-sharing}. The pilot index of the $k$-$th$ \ac{UE} is denoted by $t_k$, chosen from $t_k\in\{1,2,\ldots,\tau_p\}$. Additionally, $\mathcal{V}_k$ is a group of \ac{UEs} that share the same pilots, including the $k$-$th$ \ac{UE}. When \ac{UEs} in $\mathcal{V}_k$ transmit pilot $\boldsymbol\phi_{t_k}$, the received signal $\mathbf{y}_{t_k {l}}^\mathrm{pilot}\in\mathbb{C}^{N \times 1}$ at the $l$-$th$ \ac{AP} is as \cite{article}
%%%%%%%%%%%%%%%%%%%%%%%%%%%%%%%%%%%%%%%%%%%%%%%%%%%%%%%%%%%%%%%%%%%%%%%%%%%%%%%%%
\begin{equation} \label{co}
\mathbf{y}_{t_{k}l}^{\mathrm{pilot}}=\sum_{i\in\mathcal{V}_{k}}\sqrt{\tau_{p}p_{i}^p}~\mathbf{h}_{il}+\mathbf{n}_{t_{k}l},
\end{equation}
%%%%%%%%%%%%%%%%%%%%%%%%%%%%%%%%%%%%%%%%%%%%%%%%%%%%%%%%%%%%%%%%%%%%%%%%%%%%%%%%%
where $p_i^p$ is the pilot transmit power of \ac{UE} $i$ and the thermal noise with variance $\sigma^2$ is denoted by $\mathbf{n}_{t_kl}~\sim~\mathcal{N}_{\mathbb{C}}\left(0,\sigma^{2}\mathbf{I}_{N}\right)$. Moreover, (\ref{co}) indicates the mutual interference that is caused by the reuse of the pilot $t_k$ among the \ac{UEs} in the set $\mathcal{V}_k$ leading to pilot contamination \cite{10216314}. The \ac{MMSE} estimate of $\mathbf{h}_{kl}$ for $k \in \mathcal{V}_{k}$ can be written as
%%%%%%%%%%%%%%%%%%%%%%%%%%%%%%%%%%%%%%%%%%%%%%%%%%%%%%%%%%%%%%%%%%%%%%%%%%%%%%%%%
\begin{equation}
\hat{\mathbf{h}}_{kl}=\sqrt{\tau_{p}p_{k}^p}~\mathbf{R}_{kl}~\mathbf{\Phi}_{t_{k}l}^{-1}~\mathbf{y}_{t_{k}l}^{\mathrm{pilot}}, 
\end{equation}
%%%%%%%%%%%%%%%%%%%%%%%%%%%%%%%%%%%%%%%%%%%%%%%%%%%%%%%%%%%%%%%%%%%%%%%%%%%%%%%%%
where
%%%%%%%%%%%%%%%%%%%%%%%%%%%%%%%%%%%%%%%%%%%%%%%%%%%%%%%%%%%%%%%%%%%%%%%%%%%%%%%%%
\begin{align}
   &\mathbf{\Phi}_{t_{k}l} = \mathbb{E}\left\{\mathbf{y}_{t_{k}l}^{\mathrm{pilot}}\left(\mathbf{y}_{t_{k}l}^{\mathrm{pilot}}\right)^{\mathrm{H}}\right\} \notag \\
   &= \mathbb{E}\left\{\left(\sqrt{\tau_{p} p_{i}^p}~ \mathbf{R}_{kl} \mathbf{h}_{kl} + \mathbf{n}_{kl}\right)\left(\sqrt{\tau_{p} p_{i}^p}~ \mathbf{R}_{kl} \mathbf{h}_{kl} + \mathbf{n}_{kl}\right)^{\mathrm{H}}\right\} \notag \\
   &= \mathbb{E}\left\{\sqrt{\tau_{p} p_{i}^p}~ \mathbf{R}_{kl} \mathbf{h}_{kl} \mathbf{h}_{kl}^{\mathrm{H}} \mathbf{R}_{kl}^{\mathrm{H}} \sqrt{\tau_{p} p_{i}^p} + \mathbf{n}_{kl} \mathbf{n}_{kl}^{\mathrm{H}}\right\} \notag \\
   &= \tau_{p} p_{i}^p \mathbf{R}_{kl} \mathbb{E}\left\{\mathbf{h}_{kl} \mathbf{h}_{kl}^{\mathrm{H}}\right\} \mathbf{R}_{kl}^{\mathrm{H}} + \mathbb{E}\left\{\mathbf{n}_{kl} \mathbf{n}_{kl}^{\mathrm{H}}\right\} = \tau_{p} p_{i}^p \mathbf{R}_{kl} \mathbf{R}_{kl}^{\mathrm{H}} \notag \\
   & ~~~+ \sigma^2 \mathbf{I}_N = \sum_{i \in \mathcal{V}_{k}} \tau_{p} p_{i}^p ~\mathbf{R}_{il} + \sigma^2 \mathbf{I}_{N}.
\end{align}
%%%%%%%%%%%%%%%%%%%%%%%%%%%%%%%%%%%%%%%%%%%%%%%%%%%%%%%%%%%%%%%%%%%%%%%%%%%%%%%%%
is the correlation matrix of $\mathbf{y}_{t_{k}l}^{\mathrm{pilot}}$, having contributions from all the \ac{UEs} with same pilot sequence. The estimated channel $\hat{\mathbf{h}}_{kl}$ and the error in the estimation is $\tilde{\mathbf{h}}_{kl}=\mathbf{h}_{kl}-\hat{\mathbf{h}}_{kl}$ are independent vectors distributed as $\hat{\mathbf{h}}_{kl}\sim\mathcal{N}_{\mathbb{C}}\left(\mathbf{0},\mathbf{B}_{kl}\right)$ and $\tilde{\mathbf{h}}_{kl}\sim{\mathcal{N}}_{\mathbb{C}}\left(\mathbf{0},\mathbf{C}_{kl}\right)$, where
%%%%%%%%%%%%%%%%%%%%%%%%%%%%%%%%%%%%%%%%%%%%%%%%%%%%%%%%%%%%%%%%%%%%%%%%%%%%%%%%%
\begin{equation}
    \mathbf{B}_{kl}=\mathbb{E}\left\{\hat{\mathbf{h}}_{kl}\hat{\mathbf{h}}_{kl}^{\mathrm{H}}\right\}=\tau_{p}p_{k}^p\mathbf{R}_{kl}\mathbf{\Phi}_{t_{k}l}^{-1}\mathbf{R}_{kl},
\end{equation}
%%%%%%%%%%%%%%%%%%%%%%%%%%%%%%%%%%%%%%%%%%%%%%%%%%%%%%%%%%%%%%%%%%%%%%%%%%%%%%%%%
\begin{equation}
\mathbf{C}_{kl}=\mathbb{E}\left\{\tilde{\mathbf{h}}_{kl}\tilde{\mathbf{h}}_{kl}^{\mathrm{H}}\right\}=\mathbf{R}_{kl}-\mathbf{B}_{kl}.
\end{equation}
%%%%%%%%%%%%%%%%%%%%%%%%%%%%%%%%%%%%%%%%%%%%%%%%%%%%%%%%%%%%%%%%%%%%%%%%%%%%%%%%%
Note we assume that the network has already addressed the pilot contamination issue, as this is not the focus of our work.
%%%%%%%%%%%%%%%%%%%%%%%%%%%%%%%%%%%%%%%%%%%%%%%%%%%%%%%%%%%%%%%%%%%%%%%%%%%%%%%%%
\vspace{-5mm}
\subsection{Downlink Data Transmission and Target Sensing} \label{dlsec}
\par The \ac{DL} transmission begins with \ac{APs} simultaneously transmitting data and sensing the \ac{UE}. Using the $\hat{\mathbf{h}}_{kl}$, the \ac{APs} perform conjugate beamforming for \ac{DFRC} signal transmission to the $k$-$th$ \ac{UE}. The transmit signal for the $l$-$th$ \ac{AP} in $\mathcal{M}_k$ is
%%%%%%%%%%%%%%%%%%%%%%%%%%%%%%%%%%%%%%%%%%%%%%%%%%%%%%%%%%%%%%%%%%%%%%%%%%%%%%%%%
\begin{equation} \label{all}
\mathbf{x}_l=\sqrt{p_l}\sum_{k\in\mathcal{K}}\mathbf{w}_{kl}s_k,\:\forall l\in\mathcal{M}_k,   
\end{equation}
%%%%%%%%%%%%%%%%%%%%%%%%%%%%%%%%%%%%%%%%%%%%%%%%%%%%%%%%%%%%%%%%%%%%%%%%%%%%%%%%%
where $p_{l}$ is the normalized signal power in the \ac{DL} phase, $s_k\in\mathbb{C}$ is the data symbol for the $k$-$th$ \ac{UE} with $\mathbb{E} \{ | s_k| ^2\} = 1$, and $\mathbf{w}_{kl}\in \mathbb{C} ^{N}$ is the beamforming vector. The normalized power budget of the $l$-$th$ \ac{AP} is $\mathbb{E} \left \{ | | \mathrm{x}_l| | ^2\right \} \leq p_l$, thus
%%%%%%%%%%%%%%%%%%%%%%%%%%%%%%%%%%%%%%%%%%%%%%%%%%%%%%%%%%%%%%%%%%%%%%%%%%%%%%%%%
\begin{equation}
\sum_{k\in\mathcal{K}}\mathbf{w}_{kl}^H\mathbf{w}_{kl}\leq1,\:\forall l\in\mathcal{M}_k.
\end{equation}
%%%%%%%%%%%%%%%%%%%%%%%%%%%%%%%%%%%%%%%%%%%%%%%%%%%%%%%%%%%%%%%%%%%%%%%%%%%%%%%%%
The received superposed signal from all the \ac{APs} at the \ac{UE} is
%%%%%%%%%%%%%%%%%%%%%%%%%%%%%%%%%%%%%%%%%%%%%%%%%%%%%%%%%%%%%%%%%%%%%%%%%%%%%%%%%
\begin{equation} \mathbf{y}_{k}^{\mathrm{DL}}=\sum\limits_{l=1}^{L}\mathbf{h}_{kl}^{\mathrm{H}}\sum\limits_{i=1}^{K}\mathbf{w}_{il}\mathbf{x}_{l}+n_{k} 
=\mathbf{h}_{k}^{\mathrm{H}}\sum\limits_{i=1}^{K}\mathbf{w}_{i}\mathbf{x}_{l}+n_{k},
\end{equation}
%%%%%%%%%%%%%%%%%%%%%%%%%%%%%%%%%%%%%%%%%%%%%%%%%%%%%%%%%%%%%%%%%%%%%%%%%%%%%%%%%
where $\mathbf{w}_k$ = $\begin{bmatrix}\mathbf{w}_{k1}^{\mathrm{T}}&\ldots\mathbf{w}_{kL}^{\mathrm{T}}\end{bmatrix}^{\mathrm{T}}\in\mathbb{C}^{M}$ is the collective precoding vector, and $n_{k}\sim\mathcal{N}_{\mathbb{C}}(0,\sigma^{2})$ is the receiver noise. The most popular choice is \ac{MR} precoding with \cite{bjornson2020scalable}
%%%%%%%%%%%%%%%%%%%%%%%%%%%%%%%%%%%%%%%%%%%%%%%%%%%%%%%%%%%%%%%%%%%%%%%%%%%%%%%%%
\begin{equation}  \mathbf{w}_{il}=\sqrt{p_{i}}\frac{\widehat{\mathbf{h}}_{il}}{\sqrt{\mathbb{E}\{\|\widehat{\mathbf{h}}_{il}\|^{2}\}}},
\end{equation}
%%%%%%%%%%%%%%%%%%%%%%%%%%%%%%%%%%%%%%%%%%%%%%%%%%%%%%%%%%%%%%%%%%%%%%%%%%%%%%%%%
where $p_i\geq0$ is the transmit power allocated to the \ac{UE} $i$. \textit{\textbf{Note:}} In beamforming towards the \ac{UE}, we start with a coarse location, $\mathbf{u_k}$, assuming the \ac{UE} is within the transmission's beam width. For sensing, however, centimeter-level accuracy is often required \cite{10415170}. Therefore, we refine the \ac{UEs} position later in Section \ref{aoaclutter}.
%%%%%%%%%%%%%%%%%%%%%%%%%%%%%%%%%%%%%%%%%%%%%%%%%%%%%%%%%%%%%%%%%%%%%%%%%%%%%%%%%
\subsection{Uplink Radar Echoes Reception}
\par The \ac{APs} which transmit \ac{DL} \ac{DFRC} signal, are considered to receive radar echoes reflected from the \ac{UE} and also unwanted signals from clutter, which is considered as interference for sensing. The goal is to establish \ac{LoS} links to eliminate sensing interference from clutter, as further explained in this paper. The echo received at the $l$-$th$ \ac{AP} is
%%%%%%%%%%%%%%%%%%%%%%%%%%%%%%%%%%%%%%%%%%%%%%%%%%%%%%%%%%%%%%%%%%%%%%%%%%%%%%%%%
\begin{equation}
\begin{aligned} 
\mathbf{y}_l^{\mathrm{echo}}&=\sum_{k=1}^{N}\underbrace{\ \xi_{kl}\sqrt{\zeta_{kl}}\mathbf{a}(\phi_{0,l},\theta_{0,l})\mathbf{a}^T(\varphi_{0,k},\vartheta_{0,k})\mathbf{x}_l}_{\text{desired reflection from the UE}} \\& +\sum_{k=1}^{N}\underbrace{\mathbf{H}_{kl}\mathbf{x}_{k}}_{\text{undesired echoes from clutter}}+\mathbf{n}_{l},
\end{aligned}
\end{equation}
%%%%%%%%%%%%%%%%%%%%%%%%%%%%%%%%%%%%%%%%%%%%%%%%%%%%%%%%%%%%%%%%%%%%%%%%%%%%%%%%%
where $\mathbf{n}_{l}\sim\mathcal{CN}(\mathbf{0},\sigma^{2}\mathbf{I}_{N})$ is the receiver noise, $\zeta_{kl}$ is the channel gain including the path loss from the $l$-$th$ \ac{AP} and the $k$-$th$ \ac{UE}, while $\xi_{kl}$ is the normalized \ac{RCS} of the \ac{UE} for the respective path. Moreover, the respective array response vector is $\mathbf{a}(\phi,\theta)=\begin{bmatrix}1~ e^{j\pi\sin(\phi)\cos(\theta)} \cdots e^{j(N-1)\pi\sin(\phi)\cos(\theta)}\end{bmatrix}^T,$ where $\phi$ and $\theta$ are the azimuth and elevation angles from the \ac{AP} to the \ac{UE} respectively \cite{bjornson2017massive}, while $\mathbf{H}_{kl}$ is the target-free channel matrix between the $l$-$th$ \ac{AP} and the $k$-$th$ \ac{UE}.
%%%%%%%%%%%%%%%%%%%%%%%%%%%%%%%%%%%%%%%%%%%%%%%%%%%%%%%%%%%%%%%%%%%%%%%%%%%%%%%%%
\vspace{-3mm}
\subsection{Uplink Data Transmission}
\par The received \ac{UL} signal $\mathbf{y}_l^\mathrm{UL}\in\mathbb{C}^N$
at the $l$-$th$ \ac{AP} is
%%%%%%%%%%%%%%%%%%%%%%%%%%%%%%%%%%%%%%%%%%%%%%%%%%%%%%%%%%%%%%%%%%%%%%%%%%%%%%%%%
\begin{equation}
\mathbf{y}_l^\mathrm{UL}=\sum_{i=1}^K\mathbf{h}_{il}s_i+\mathrm{n}_l,    
\end{equation}
%%%%%%%%%%%%%%%%%%%%%%%%%%%%%%%%%%%%%%%%%%%%%%%%%%%%%%%%%%%%%%%%%%%%%%%%%%%%%%%%%
where $s_i \sim \mathcal{N}\mathbb{C}(0, p_i)$ represents the signal transmitted by the \ac{UE} $i$ with power $p_i$, and $n_l \sim \mathcal{N}\mathbb{C}(0, \sigma^2 \mathbf{I}_N)$ is the noise. Each \ac{AP} estimates the data locally before forwarding it to the \ac{CPU} for final decoding, as illustrated in Fig. \ref{Signal}. The combining vector used by the $l$-$th$ \ac{AP} for the $k$-$th$ \ac{UE} is denoted as $\mathbf{a}_{kl} \in \mathbb{C}^N$, where $k \in \mathcal{D}_l$. The local estimate of $s_k$ can then be expressed as
%%%%%%%%%%%%%%%%%%%%%%%%%%%%%%%%%%%%%%%%%%%%%%%%%%%%%%%%%%%%%%%%%%%%%%%%%%%%%%%%%
\begin{equation} \label{ull}
\begin{aligned}
\tilde{s}_{kl}&=\mathbf{a}_{kl}^\mathrm{H}\mathbf{D}_{kl}\mathbf{y}_l^\mathrm{UL}\\&=\mathbf{a}_{kl}^\mathrm{H}\mathbf{D}_{kl}\mathbf{h}_{kl}s_{k}+\mathrm{a}_{kl}^\mathrm{H}\mathbf{D}_{kl}\sum_{i=1,\:i\neq k}^K\mathbf{h}_{il}s_i+\mathbf{a}_{kl}^\mathrm{H}\mathbf{D}_{kl}\mathbf{n}_l.
\end{aligned}
\end{equation}
%%%%%%%%%%%%%%%%%%%%%%%%%%%%%%%%%%%%%%%%%%%%%%%%%%%%%%%%%%%%%%%%%%%%%%%%%%%%%%%%%
Any combining vector can be utilized in the above expression. \ac{MR} combining, with $\mathbf{a}_{kl}^{\mathrm{MR}} = \hat{\mathbf{h}}_{kl}$, was employed in \cite{7869024}. More recently, \cite{bjornson2020scalable} proposed the use of local partial \ac{MMSE} combining
%%%%%%%%%%%%%%%%%%%%%%%%%%%%%%%%%%%%%%%%%%%%%%%%%%%%%%%%%%%%%%%%%%%%%%%%%%%%%%%%%
\begin{equation}
\mathbf{a}_{kl}^{\mathrm{LP-MMSE}}=p_k\left(\sum_{i\in\mathcal{D}_l}p_i\left(\hat{\mathbf{h}}_{il}\hat{\mathbf{h}}_{il}^\mathrm{H}+\mathbf{C}_{il}\right)+\sigma^2\mathrm{I}_N\right)^{-1}\hat{\mathbf{h}}_{kl}.
\end{equation}
%%%%%%%%%%%%%%%%%%%%%%%%%%%%%%%%%%%%%%%%%%%%%%%%%%%%%%%%%%%%%%%%%%%%%%%%%%%%%%%%%
The local estimates $\{\tilde{s}_{kl}\}$ are subsequently transmitted to the \ac{CPU}, where they are linearly combined with weights $\{w_{kl}\}$ to produce $\hat{s}_k=\sum_{l=1}^{L}w_{kl}^{*}\tilde{s}_{kl}$, which is then utilized for decoding $s_k$. Using (\ref{ull}), the final estimate of $s_k$ is derived as follows
%%%%%%%%%%%%%%%%%%%%%%%%%%%%%%%%%%%%%%%%%%%%%%%%%%%%%%%%%%%%%%%%%%%%%%%%%%%%%%%%%
\begin{equation}
\hat{s}_k=\mathbf{a}_k^\mathrm{H}\mathbf{W}_k^\mathrm{H}\mathbf{D}_k\mathbf{h}_ks_k+\sum_{i=1,\:i\neq k}^K\mathbf{a}_k^\mathrm{H}\mathbf{W}_k^\mathrm{H}\mathbf{D}_k\mathbf{h}_is_i+\mathbf{a}_k^\mathrm{H}\mathbf{W}_k^\mathrm{H}\mathbf{D}_k\mathrm{n}, 
\end{equation}
%%%%%%%%%%%%%%%%%%%%%%%%%%%%%%%%%%%%%%%%%%%%%%%%%%%%%%%%%%%%%%%%%%%%%%%%%%%%%%%%%
where $\mathbf{W}_k=\mathrm{diag}\left(w_{k1}\mathbf{I}_N,\ldots,w_{kL}\mathbf{I}_N\right)\in\mathbb{C}^{(LN)\times(LN)}.$
%%%%%%%%%%%%%%%%%%%%%%%%%%%%%%%%%%%%%%%%%%%%%%%%%%%%%%%%%%%%%%%%%%%%%%%%%%%%%%%%%
\section{Problem Formulation}
\par In this section, we first explain the concepts of \ac{SE} and \ac{SCNR}, on which our optimization problems are based. Following this, we present the optimization problems that ensure effective \ac{JRC} \ac{UA} in \ac{CF-mMIMO} systems.
%%%%%%%%%%%%%%%%%%%%%%%%%%%%%%%%%%%%%%%%%%%%%%%%%%%%%%%%%%%%%%%%%%%%%%%%%%%%%%%%%
\subsection{Achievable Spectral Efficiency}
\par The achievable \ac{SE} for the $k$-$th$ \ac{UE} in \ac{CF-mMIMO} network is expressed as \cite{bjornson2020scalable}
%%%%%%%%%%%%%%%%%%%%%%%%%%%%%%%%%%%%%%%%%%%%%%%%%%%%%%%%%%%%%%%%%%%%%%%%%%%%%%%%%
\begin{equation}
 \mathrm{SE}_k=\left(1-\frac{\tau_p}{\tau_c}\right)\log_2(1+\mathrm{SINR}_k),  
\end{equation}
%%%%%%%%%%%%%%%%%%%%%%%%%%%%%%%%%%%%%%%%%%%%%%%%%%%%%%%%%%%%%%%%%%%%%%%%%%%%%%%%%
where the effective \ac{SINR} of the $k$-$th$ \ac{UE} is denoted as $\mathrm{SINR}_k$ . In (\ref{bottom}), 
%%%%%%%%%%%%%%%%%%%%%%%%%%%%%%%%%%%%%%%%%%%%%%%%%%%%%%%%%%%%%%%%%%%%%%%%%%%%%%%%%
\begin{figure*}[ht]
\begin{equation} \label{bottom}
\begin{aligned}
\mathrm{SINR}_{k}& =\frac{p_k\big|\mathbb{E}\left\{\mathbf{a}_k^\mathrm{H}\mathbf{W}_k^\mathrm{H}\mathbf{D}_k\mathbf{h}_k\right\}\big|^2}{\sum_{i=1}^Kp_i \mathbb{E}\left\{\left|\mathbf{a}_k^\mathrm{H}\mathbf{W}_k^\mathrm{H}\mathbf{D}_k\mathbf{h}_i\right|^2\right\}-p_k\left|\mathbb{E}\left\{\mathbf{a}_k^\mathrm{H}\mathbf{W}_k^\mathrm{H}\mathbf{D}_k\mathbf{h}_k\right\}\right|^2+\sigma^2 \mathbb{E}\left\{\left\|\mathbf{D}_k\mathbf{W}_k^\mathrm{H}\mathbf{a}_k\right\|^2\right\}} \\
\mathrm{SINR}_{k}&=\frac{p_{k}\big|\mathbf{w}_{k}^{\mathrm{H}}\mathbf{v}_{k}\big|^{2}}{\mathbf{w}_{k}^{\mathrm{H}}\left(\sum_{i=1}^{K}p_{i}\mathbf{\Lambda}_{ki}^{(1)}-p_{k}\mathbf{v}_{k}\mathbf{v}_{k}^{\mathrm{H}}+\sigma^{2}\mathbf{\Lambda}_{k}^{(2)}\right)\mathbf{w}_{k}},
\end{aligned}
\end{equation}
\hrulefill
\end{figure*}
%%%%%%%%%%%%%%%%%%%%%%%%%%%%%%%%%%%%%%%%%%%%%%%%%%%%%%%%%%%%%%%%%%%%%%%%%%%%%%%%%
%%%%%%%%%%%%%%%%%%%%%%%%%%%%%%%%%%%%%%%%%%%%%%%%%%%%%%%%%%%%%%%%%%%%%%%%%%%%%%%%%
\begin{subequations}
\begin{align}
\mathbf{w}_{k}&=\begin{bmatrix}w_{kl},\ldots,w_{kL}\end{bmatrix}^\mathrm{T}, \\
\mathbf{v}_{k}& =\begin{bmatrix}\mathbb{E}\left\{\mathbf{a}_{k1}^\mathrm{H}\mathbf{D}_{k1}\mathbf{h}_{k1}\right\},\ldots,\mathbb{E}\left\{\mathbf{a}_{kL}^\mathrm{H}\mathbf{D}_{kL}\mathbf{h}_{kL}\right\}\end{bmatrix}^\mathrm{T}, \\
\Lambda_{ki}^{(1)}& =\begin{bmatrix}\mathbb{E}\{\mathbf{a}_{kl}^\mathrm{H}\mathbf{D}_{kl}\mathbf{h}_{il}\mathbf{h}_{ij}^\mathrm{H}\mathbf{D}_{kj}\mathbf{a}_{kj}\}:l,j=1,\ldots,L\end{bmatrix}, \\
\Lambda_{k}^{(2)}& =\mathrm{diag}\left(\mathbb{E}\left\{\left\|\mathbf{D}_{k1}\mathbf{a}_{k1}\right\|^2\right\},\ldots,\mathbb{E}\left\{\left\|\mathbf{D}_{kL}\mathbf{a}_{kL}\right\|^2\right\}\right), 
\end{align}
\end{subequations}
%%%%%%%%%%%%%%%%%%%%%%%%%%%%%%%%%%%%%%%%%%%%%%%%%%%%%%%%%%%%%%%%%%%%%%%%%%%%%%%%%
\vspace{-5mm}
\subsection{Signal-Clutter plus Noise Ratio}
\par For a specified range resolution cell of the \ac{AP} used for sensing, the \ac{SCNR} is $\frac{P_{\mathrm{ue}}}{P_{\mathrm{c}} + P_{\mathrm{n}}}$, where $P_{\mathrm{ue}}$ and $P_{\mathrm{c}}$ are the powers of the \ac{UE} and clutter backscatter, respectively, and $P_{\mathrm{n}}$ is the noise power. We consider a scenario with multiple discrete scatterers making up the clutter $\mathbf{c}$ around a specific \ac{UE}, these scatterers' positions are modeled as a 2D \ac{PPP} ($\Gamma'$) \cite{chiu2013stochastic}, with a uniform distribution. The \ac{RCS} of these scatterers follows a standard Swerling cross-section model as described in \cite{skolnik1980introduction}. The closest distance between the clutter scatterers and the \ac{AP} is $R$, which is within the radar's maximum unambiguous range. The \ac{RCS} of each scatterer $(\upsilon_{\mathbf{c}})$ contributing to the clutter follows the distribution
%%%%%%%%%%%%%%%%%%%%%%%%%%%%%%%%%%%%%%%%%%%%%%%%%%%%%%%%%%%%%%%%%%%%%%%%%%%%%%%%%
\begin{equation} \label{dis}
 P(\upsilon_{\mathbf{c}})=\frac{1}{\upsilon_{\mathrm{c_{avg}}}}\exp\left(-\frac{\upsilon_{\mathbf{c}}}{\upsilon_{\mathrm{c_{avg}}}}\right),
\end{equation}
%%%%%%%%%%%%%%%%%%%%%%%%%%%%%%%%%%%%%%%%%%%%%%%%%%%%%%%%%%%%%%%%%%%%%%%%%%%%%%%%%
where $\upsilon_{\mathrm{c_{avg}}}$ is the average clutter cross-section. For each realization of the \ac{PPP}, the total received clutter power is \cite{skolnik1980introduction}
%%%%%%%%%%%%%%%%%%%%%%%%%%%%%%%%%%%%%%%%%%%%%%%%%%%%%%%%%%%%%%%%%%%%%%%%%%%%%%%%%
\begin{equation}
\mathbf{C}=\sum_{c\in\Gamma'}\frac{ZG(\theta_\mathbf{c})\upsilon_{\mathbf{c}}g_c}{\mathbf{r_c}^{2q}},
\end{equation}
%%%%%%%%%%%%%%%%%%%%%%%%%%%%%%%%%%%%%%%%%%%%%%%%%%%%%%%%%%%%%%%%%%%%%%%%%%%%%%%%%
where $Z$ is a constant and defined as $Z = \frac{p_{l}\lambda^{2}}{(4\pi)^{3}}$, $G(\theta_{\mathbf{c}}) = G_{tx}(\theta_{\mathbf{c}})G_{rx}(\theta_{\mathbf{c}})$ ($G_{tx}(\theta)$ and $G_{rx}(\theta)$ being the directional transmit and receive antenna gains, respectively), $g_c$ is a random variable modeling the fading between clutter returns, and $q$ is path loss exponent. The position of each scatterer $\mathbf{c}$ is given in polar coordinates as $\vec{\mathbf{r}}_c = (\mathbf{r}_{\mathbf{c}}, \theta_{\mathbf{c}})$, with azimuth $\theta_\mathbf{c}$ uniformly distributed in $[0, 2\pi)$ and range $\mathbf{r_c}$ in $(R, \infty]$. For a \ac{UE} cross-section of $\upsilon_{\mathbf{t}}$, the received signal at the \ac{AP} is $\mathbf{S}_s = \frac{ZG(\theta_\mathbf{t})\upsilon_\mathbf{t}}{\mathbf{r_t}^{2q}}$, where $G(\theta_\mathbf{t}) = G_{tx}(\theta_\mathbf{t}) G_{rx}(\theta_\mathbf{t})$. We assume a Swerling-1 \ac{RCS} fluctuation for the \ac{UE}.
%%%%%%%%%%%%%%%%%%%%%%%%%%%%%%%%%%%%%%%%%%%%%%%%%%%%%%%%%%%%%%%%%%%%%%%%%%%%%%%%%
\begin{equation} \label{swerling}
P(\upsilon_{\mathbf{t}})=\frac{1}{\upsilon_{\mathrm{t_{avg}}}}\exp\left(-\frac{\upsilon_{\mathbf{t}}}{\upsilon_{\mathrm{t_{avg}}}}\right),
\end{equation}
%%%%%%%%%%%%%%%%%%%%%%%%%%%%%%%%%%%%%%%%%%%%%%%%%%%%%%%%%%%%%%%%%%%%%%%%%%%%%%%%%
where $\upsilon_{\mathrm{t_{avg}}}$ represents the average \ac{UE} cross-section. Under \ac{LoS} conditions, the minimum \ac{SCNR} occurs when $\mathbf{r_t}=R$. Assuming the \ac{AP} tracks the \ac{UE} within its main beam, such that $G(\theta_t)=1$ therefore, the average \ac{SCNR} for a given $R$ is
%%%%%%%%%%%%%%%%%%%%%%%%%%%%%%%%%%%%%%%%%%%%%%%%%%%%%%%%%%%%%%%%%%%%%%%%%%%%%%%%%
\begin{equation}
\mathbb{E}\left[\mathbf{SCNR}(R)\right]=\mathbb{E}_{\upsilon_{t},\upsilon_{c},g_{c},\Gamma^{\prime}}\left[\frac{\frac{Z\upsilon_{\mathbf{t}}}{R^{2q}}}{\mathrm{n}_{l}+\sum_{\mathbf{c}\in\Gamma}\frac{ZG(\theta_{\mathbf{c}})\upsilon_{\mathbf{c}}g_c}{\mathbf{r_c}^{2q}}}\right].
\end{equation}
%%%%%%%%%%%%%%%%%%%%%%%%%%%%%%%%%%%%%%%%%%%%%%%%%%%%%%%%%%%%%%%%%%%%%%%%%%%%%%%%% 
\par In the context of the proposed \ac{JRC} \ac{CF-mMIMO} scenario, mitigating clutter is crucial for accurate \ac{UE} detection/tracking and \ac{AP} association. One important metric in understanding the impact of clutter in front of a specific \ac{UE} and \ac{AP} view angle, is the \ac{$P_{dc}$}. This is defined as the probability that the average \ac{SCNR} at a distance $r$ is greater than or equal to a predefined threshold $\gamma$
%%%%%%%%%%%%%%%%%%%%%%%%%%%%%%%%%%%%%%%%%%%%%%%%%%%%%%%%%%%%%%%%%%%%%%%%%%%%%%%%%
\begin{equation}
P_{dc}(r)\triangleq\mathbb{P}\left(\mathbf{SCNR}(r)\geq\gamma\right).
\end{equation}
%%%%%%%%%%%%%%%%%%%%%%%%%%%%%%%%%%%%%%%%%%%%%%%%%%%%%%%%%%%%%%%%%%%%%%%%%%%%%%%%%
In the following theorems, we will delve deeper into the mathematical analysis to characterize \ac{$P_{dc}$} in both \ac{LoS} and \ac{NLoS} conditions, highlighting the significance of the proposed \ac{UA}. These analyses will provide better insight into the clutter surrounding a \ac{UE} and underscore the importance of considering clutter factor when selecting \ac{APs}.

\textbf{Theorem 1.} \textit{The \ac{$P_{dc}$} for the \ac{UE} located at \ac{LoS} distance $R$ from a \ac{AP} without any clutter, such that $G(\theta _{t})$ = 1 is as}
%%%%%%%%%%%%%%%%%%%%%%%%%%%%%%%%%%%%%%%%%%%%%%%%%%%%%%%%%%%%%%%%%%%%%%%%%%%%%%%%%
\begin{equation} \label{pdcr}
P_{dc}(R)=I(R)exp\left(\frac{-\gamma \mathrm{n}_{l}R^{2q}}{Z\:\upsilon_{\mathrm{t_{avg}}}}\right),
\end{equation}
%%%%%%%%%%%%%%%%%%%%%%%%%%%%%%%%%%%%%%%%%%%%%%%%%%%%%%%%%%%%%%%%%%%%%%%%%%%%%%%%%
\textrm{where} $I(R)=exp\left(-\varrho\int_{0}^{2\pi}\int_{R}^{R+\Delta R}\frac{\nu(R)G(\theta_{c})r_{c}}{\nu(R)G(\theta_{c})+r_{c}^{2q}}dr_{c}d\theta_{c}\right),$ \textrm{and} $\nu(R)=\frac{\gamma R^{2q}\upsilon_{\mathrm{c_{avg}}}}{\upsilon_{\mathrm{t_{avg}}}}$, From (\ref{pdcr}) we have,

%%%%%%%%%%%%%%%%%%%%%%%%%%%%%%%%%%%%%%%%%%%%%%%%%%%%%%%%%%%%%%%%%%%%%%%%%%%%%%%%%
{\begin{subequations} \label{der}
{\small
\begin{align}
&P_{dc}(R)=\mathbb{P}\left[\upsilon_{t}>\frac{\gamma \mathrm{n}_{l}R^{2q}}{Z}+\gamma R^{2q}\sum_{\mathrm{c}\in\Gamma}\frac{G(\theta_{\mathrm{c}})\upsilon_{\mathrm{c}}\mathbf{g}_{\mathrm{c}}}{\mathbf{r}_{\mathrm{c}}^{2q}}\right]\\
&=\mathbb{E}_{\upsilon_{c},g_{c},\Gamma^{\prime}}\left[exp\left(\frac{-\gamma \mathrm{n}_{l}R^{2q}}{Z\:\upsilon_{\mathrm{t_{avg}}}}-\frac{\gamma R^{2q}}{\upsilon_{\mathrm{t_{avg}}}}\sum_{\mathrm{c}\in\Gamma}\frac{G(\theta_{\mathrm{c}})\upsilon_{\mathrm{c}}\mathbf{g}_{\mathrm{c}}}{\mathbf{r}_{\mathrm{c}}^{2q}}\right)\right]\\
&=exp\left(\frac{-\gamma \mathrm{n}_{l}R^{2q}}{Z\upsilon_{t_{avg}}}\right)\mathbb{E}_{\upsilon_{c},g_{c},\Gamma^{\prime}}\left[\prod_{c\in\Gamma}exp\left(\frac{-\gamma R^{2q}G(\theta_{\mathrm{c}})\upsilon_{\mathrm{c}}\mathbf{g}_{\mathrm{c}}}{\upsilon_{\mathrm{t_{avg}}}\:\mathbf{r}_{\mathrm{c}}^{2q}}\right)\right],
\end{align}}
\end{subequations}
}
where (\ref{der}a) assumes \ac{UE} cross-section follows (\ref{swerling}), while (\ref{der}b) results from taking the expectation, considering clutter scatterers, cross-sections, spatial distribution, and mutual interference. Using the property that exponentials of sums equal products of exponential functions yields (\ref{der}c). Additionally, evaluating the expectation using the probability generating functional of the point process $\Gamma$ gives \cite{chiu2013stochastic} 
%%%%%%%%%%%%%%%%%%%%%%%%%%%%%%%%%%%%%%%%%%%%%%%%%%%%%%%%%%%%%%%%%%%%%%%%%%%%%%%%%

{\small
\begin{equation} \label{subi}
\begin{aligned}
&\mathbb{E}_{\upsilon_{c},g_{c},\Gamma^{\prime}}\left[\prod_{c\in\Gamma^{\prime}} exp\left(\frac{-\gamma R^{2q}G(\theta_{\mathrm{c}})\upsilon_{\mathrm{c}}\mathrm{g_{\mathrm{c}}}}{\upsilon_{\mathrm{t_{avg}}}\mathrm{r_{c}}^{2q}}\right)\right]=\\
& exp\left(-\varrho\int_{\mathbb{R}^{2}}\left(1-\mathbb{E}_{\upsilon_{\mathrm{c}},\mathrm{g_{c}}}\left[exp\left(-\frac{\gamma R^{2q}G(\theta_{c})\upsilon_{\mathrm{c}}\mathrm{g_{c}}}{\upsilon_{\mathrm{t_{avg}}}r_{c}^{2q}}\right)\right]d(\vec{r_{c}})\right)\right).
\end{aligned}
\end{equation}
}

%%%%%%%%%%%%%%%%%%%%%%%%%%%%%%%%%%%%%%%%%%%%%%%%%%%%%%%%%%%%%%%%%%%%%%%%%%%%%%%%%
\par The expectation inside the integral is taken for the interference $g_c$ and the scatterer cross-section $\upsilon_c$. Under the worst scenario, $g_c = 1$ (indicating that signals from all scatterers add constructively, resulting in the maximum clutter returns), and assuming the scatterer cross-section follows (\ref{dis}), we derive
%%%%%%%%%%%%%%%%%%%%%%%%%%%%%%%%%%%%%%%%%%%%%%%%%%%%%%%%%%%%%%%%%%%%%%%%%%%%%%%%%
\begin{equation}
\begin{aligned}
&\mathbb{E}_{\upsilon_{\mathbf{e}},\mathbf{g_{c}}}\left[exp\left(-\frac{\gamma R^{2q}G(\theta_{c})\upsilon_{\mathbf{c}}\mathbf{g_{c}}}{\upsilon_{\mathrm{t_{avg}}}r_{c}^{2q}}\right)\right]=\frac{1}{1+\frac{\gamma R^{2q}G(\theta_{c})\:\upsilon_{\mathrm{c_{avg}}}}{\upsilon_{\mathrm{t_{avg}}}r_{c}^{2q}}}\\&=\frac{1}{1+\frac{\nu(R)G(\theta_{c})}{r_{c}^{2q}}},
\end{aligned}
\end{equation}
%%%%%%%%%%%%%%%%%%%%%%%%%%%%%%%%%%%%%%%%%%%%%%%%%%%%%%%%%%%%%%%%%%%%%%%%%%%%%%%%%
where $\nu(R)=\frac{\gamma R^{2q}\upsilon_{\mathrm{c_{avg}}}}{\upsilon_{\mathrm{tavg}}}.$ Now, substituting this into (\ref{subi})
%%%%%%%%%%%%%%%%%%%%%%%%%%%%%%%%%%%%%%%%%%%%%%%%%%%%%%%%%%%%%%%%%%%%%%%%%%%%%%%%%
\begin{equation}
\begin{aligned}
&\mathbb{E}_{\upsilon_{c},g_{c},\Gamma^{\prime}}\left[\prod_{c\in\Gamma^{\prime}} \exp\left( -\frac{\gamma R^{2q} G(\theta_c) \upsilon_c g_c}{\upsilon_{{t_{avg}}} r_c^{2q}} \right)\right] \\
&\quad= \exp\left(-\varrho \int_{\mathbb{R}^{2}} \left(1 - \frac{1}{1 + \frac{\nu(R) G(\theta_c)}{r_c^{2q}}} \right) d(\vec{r_c}) \right),
\end{aligned}
\end{equation}
%%%%%%%%%%%%%%%%%%%%%%%%%%%%%%%%%%%%%%%%%%%%%%%%%%%%%%%%%%%%%%%%%%%%%%%%%%%%%%%%%
Simplifying inside the exponential
%%%%%%%%%%%%%%%%%%%%%%%%%%%%%%%%%%%%%%%%%%%%%%%%%%%%%%%%%%%%%%%%%%%%%%%%%%%%%%%%%
\begin{equation}
    1 - \frac{1}{1 + \frac{\nu(R) G(\theta_c)}{r_c^{2q}}} = \frac{\frac{\nu(R) G(\theta_c)}{r_c^{2q}}}{1 + \frac{\nu(R) G(\theta_c)}{r_c^{2q}}}= \frac{\nu(R) G(\theta_c) r_c^{2q}}{\nu(R) G(\theta_c) + r_c^{2q}},
\end{equation}
%%%%%%%%%%%%%%%%%%%%%%%%%%%%%%%%%%%%%%%%%%%%%%%%%%%%%%%%%%%%%%%%%%%%%%%%%%%%%%%%%
So the integral becomes
%%%%%%%%%%%%%%%%%%%%%%%%%%%%%%%%%%%%%%%%%%%%%%%%%%%%%%%%%%%%%%%%%%%%%%%%%%%%%%%%%
\begin{equation} \label{ir}
\exp\left(-\varrho \int_{0}^{2\pi} \int_{R}^{R+\Delta R} \frac{\nu(R) G(\theta_c) r_c}{\nu(R) G(\theta_c) + r_c^{2q}} dr_c d\theta_c \right),
\end{equation}
%%%%%%%%%%%%%%%%%%%%%%%%%%%%%%%%%%%%%%%%%%%%%%%%%%%%%%%%%%%%%%%%%%%%%%%%%%%%%%%%%
where (\ref{ir}) is the $I(R)$, which was defined in \textbf{Theorem 1}. Note that in this scenario clutter from scatterers that lie within the same range resolution cell of the \ac{UE} is considered ($R$ and $R+\Delta R$, where $\Delta R=c/2BW$, the range resolution is estimated from the bandwidth, $BW$ of the radar).
\par Furthermore, when considering \ac{NLoS} scenario where clutter obstructs the view angle between the \ac{AP} and the \ac{UE}, the radar signal to the \ac{UE} and its return to the \ac{AP} undergo exponential decay due to propagation through the material of the clutter scatterers. The \ac{UE} can be detected when the \ac{SCNR} at the range cell occupied by the \ac{UE} meets or exceeds the predefined threshold. The clutter consists of a set of discrete scatterers, whose positions $(\vec{r_c}=\mathbf{r_c},\theta_{\mathbf{c}})$ follow a \ac{PPP} $\Gamma$ defined in $\mathbb{R}^2$, with \ac{UE} located at $\vec{r}_t=(r_t,\theta_t)$. The attenuation factor due to the material properties of the scatterer is assumed to be $\alpha(\lambda_{\mathrm{c}})$. Since the region between the radar and the target is partially covered with clutter, the attenuation $\alpha$ is factored with $\varrho\upsilon_0$ where $\varrho$ is the intensity measure of the \ac{PPP} while $\upsilon_0$ is the average physical area occupied by the scatterers. Then the received signal from the \ac{UE} is $\mathbf{S}=\frac{ZG(\theta_t)\upsilon_\mathbf{t}e^{-2\alpha'r_t}}{r_t^{2q}}$. Similarly, the clutter returns are
\begin{equation}
\mathrm{C}=\sum_{\mathrm{c}\in\Gamma}\frac{ZG(\theta_{\mathbf{c}})e^{-2\alpha^{\prime}\mathbf{r_{c}}}\upsilon_{\mathbf{c}}g_{c}}{\mathrm{r_{c}}^{2q}}.
\end{equation}
\par \textbf{Theorem 2.} \textit{The \ac{$P_{dc}$} of a \ac{UE} at \ac{NLoS} distance $r_t$ from the \ac{AP} within the field of view such that $G\left(\theta_t\right)=1$ is}
\begin{equation}
    P_{D C}\left(r_t\right)=J\left(r_t\right) \exp \left(\frac{-\gamma \mathrm{n}_{l} r_t^{2 q} e^{2 \alpha^{\prime} r_t}}{Z \upsilon_{\mathrm{t_{avg}}}}\right)
\end{equation}
where
\begin{equation}
\begin{aligned}
& J\left(r_t\right)= \\
& \exp \left(-\varrho \int_0^{2 \pi} \int_{r_t}^{r_t+\Delta R} \frac{\nu^{\prime}(r) G\left(\theta_c\right) r_c}{\nu^{\prime}(r) G\left(\theta_c\right)+r_c^{2 q} e^{2 \alpha^{\prime} r_c}} d r_c d \theta_c\right),
\end{aligned}
\end{equation}
and
\begin{equation}
\nu^{\prime}\left(r_t\right)=\frac{\gamma r_t^{2 q} e^{2 \alpha^{\prime} r_t} \upsilon_{\mathrm{c}_{avg}}}{\upsilon_{t_{avg}}} .
\end{equation}
From the definition of \ac{$P_{dc}$}, we have (\textit{Proof.} See Appendix),
\begin{equation}
\begin{aligned}
& P_{D C}\left(r_t\right)=\mathbb{P}\left[\operatorname{SCNR}\left(r_t\right)>\gamma\right]= \\
& \mathbb{P}\left[\frac{\frac{Z \upsilon_{\mathrm{t}}}{r_t^{2 q}} e^{-2 \alpha^{\prime} r_t}}{\mathrm{n}_{l}+\sum_{\mathbf{c} \in \Gamma} \frac{Z G\left(\theta_{\mathbf{c}}\right) \upsilon_{\mathbf{c}} \mathbf{g}_{\mathrm{c}}}{\mathbf{r}_{\mathrm{c}}{ }^{2 q}}} e^{-2 \alpha^{\prime} \mathbf{r}_{\mathbf{c}}}>\gamma\right].
\end{aligned}
\end{equation}
%%%%%%%%%%%%%%%%%%%%%%%%%%%%%%%%%%%%%%%%%%%%%%%%%%%%%%%%%%%%%%%%%%%%%%%%%%%%%%%%% 
\begin{figure*}
\centering
\resizebox{2\columnwidth}{!}{
\includegraphics{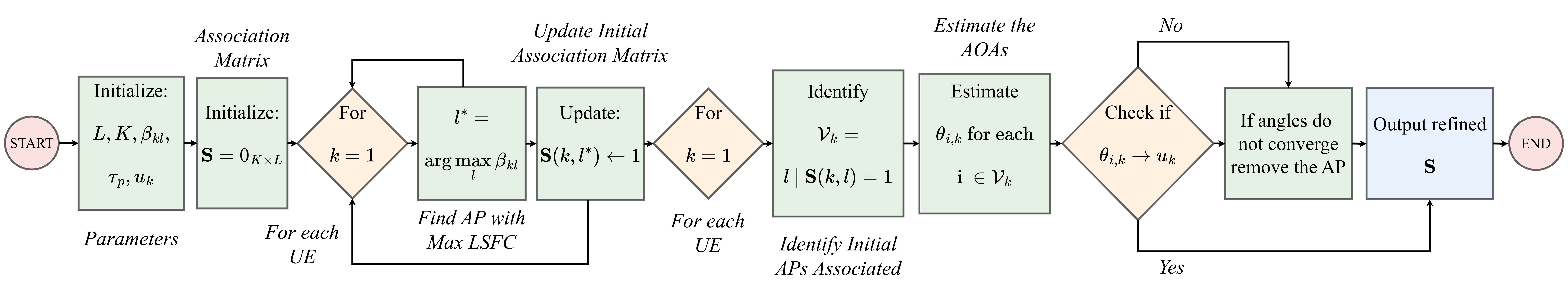}}
\caption{Flowchart of the proposed scheme.}
\label{flowcc}
\end{figure*}
%%%%%%%%%%%%%%%%%%%%%%%%%%%%%%%%%%%%%%%%%%%%%%%%%%%%%%%%%%%%%%%%%%%%%%%%%%%%%%%%%
\vspace{-5mm}
\subsection{Optimization Problems}
\par When the $k$-$th$ \ac{UE} accesses the network, it selects its serving \ac{APs} from $\mathcal{M}_k$. This selection is limited because each \ac{AP} can only accommodate up to $\tau_p$ \ac{UEs}. Ensuring scalability in \ac{CF-mMIMO} systems hinges on two crucial assumptions, limiting each \ac{AP} to $\tau_p$ associated \ac{UEs} and ensuring all $N$ antennas of \ac{APs} serve the \ac{UEs}, ($\left|\mathcal{D}_l\right| \leq \tau_p$)
%%%%%%%%%%%%%%%%%%%%%%%%%%%%%%%%%%%%%%%%%%%%%%%%%%%%%%%%%%%%%%%%%%%%%%%%%%%%%%%%%
\begin{equation}
\mathbf{D}_{k l}=\left\{\begin{array}{ll}
\mathbf{I}_N & \text {if } k \in \mathcal{D}_l \\
\mathbf{0}_N & \text {otherwise}
\end{array}\right., \textrm{for}~ l=1, \ldots, L.
\end{equation}
%%%%%%%%%%%%%%%%%%%%%%%%%%%%%%%%%%%%%%%%%%%%%%%%%%%%%%%%%%%%%%%%%%%%%%%%%%%%%%%%%
%%%%%%%%%%%%%%%%%%%%%%%%%%%%%%%%%%%%%%%%%%%%%%%%%%%%%%%%%%%%%%%%%%%%%%%%%%%%%%%%%
\begin{algorithm} 
\caption{Initial Phase \ac{UA}}
\begin{algorithmic}[1]
\STATE \textbf{Input:} $L, K, \{\beta_{kl}\}, \tau_p$
\STATE \textbf{Output:} $\mathbf{S}$
\STATE Initialize: $\mathbf{S} = 0_{K \times L}$
\FOR{$k = 1:K$}
    \STATE $l^* = \arg\max_l \beta_{kl}$
    \STATE $\mathbf{S}(k, l^*) \leftarrow 1$
\ENDFOR
\FOR{$l = 1:L$}
    \STATE $\mathcal{L}_{\mathrm{UE}, l} = \mathrm{argsort}(\beta_{kl}, \text{ descending})$
    \STATE $\mathcal{B}_l = \text{first } (\tau_p - |\mathcal{M}_l|) \text{ elements of } \mathcal{L}_{\mathrm{UE}, l}$
    \FOR{each $k$ in $\mathcal{B}_l$}
        \STATE $\mathbf{S}(k, l) \leftarrow 1$
    \ENDFOR
\ENDFOR
\FOR{$l = 1:L$}
    \STATE $\mathcal{L}_{R,l} = \text{find repeated UEs in } \mathcal{M}_l$
    \WHILE{$|\mathcal{L}_{R,l}| > 0$}
        \STATE $k^* = (\tau_p - |\mathcal{M}_l| + i)$th element of $\mathcal{L}_{\mathrm{UE}, l}$, where $i = 1, 2, \ldots, |\mathcal{L}_{R,l}|$
        \STATE $\mathbf{S}(k^*, l) \leftarrow 1$
        \STATE $\mathcal{L}_{R,l} = \text{update repeated UEs in } \mathcal{M}_l$
    \ENDWHILE
\ENDFOR
\end{algorithmic}
\end{algorithm} 
%%%%%%%%%%%%%%%%%%%%%%%%%%%%%%%%%%%%%%%%%%%%%%%%%%%%%%%%%%%%%%%%%%%%%%%%%%%%%%%%%

The \ac{UA} problem on a given pilot assignment is
%%%%%%%%%%%%%%%%%%%%%%%%%%%%%%%%%%%%%%%%%%%%%%%%%%%%%%%%%%%%%%%%%%%%%%%%%%%%%%%%%
\begin{subequations} \label{P1}
\begin{align}
\mathrm{\mathbf{P1:}}&\max_{\mathbf{S}^{(j)}}\sum_{k=1}^{K}~\mathrm{SE}_{k}^{(j)}~\forall j=1,\ldots,J\\
\mathrm{s.t.}&|\mathcal{M}_{l}^{(j)}| \leq \tau_{p} ~\forall j=1,\ldots,J~\forall ~l=1,\ldots,L\\
&|\mathcal{V}_{k}^{(j)}| \geq 2~\forall j=1,\ldots,J~\forall~ k=1,\ldots,K
\end{align}
\end{subequations}
%%%%%%%%%%%%%%%%%%%%%%%%%%%%%%%%%%%%%%%%%%%%%%%%%%%%%%%%%%%%%%%%%%%%%%%%%%%%%%%%%
where $\mathbf{S}^{(j)}$ is the association matrix and $\mathrm{SE}_k^{(j)}$ is the $k$-$th$ \ac{UE} \ac{SE} under the $j$-$th$ combination, resulting in $J = \binom{K}{\tau_p}^L$, where $\binom{K}{\tau_p}$ denotes combinations of $\tau_p$ \ac{UEs} from $K$ for a single \ac{AP}.
\par The constraint (\ref{P1}b) ensures scalability, limiting number of \ac{UEs} each \ac{AP} can serve. While (\ref{P1}c) ensures each \ac{UE} is served by at least two \ac{APs}, ensuring coverage and reliability. This also facilitates angulation-based clutter detection (Section \ref{aoaclutter}), enhancing sensing accuracy. \textbf{P1} is \ac{NP}-hard due to its combinatorial nature, but optimal solutions can be explored using methods like exhaustive search across all $J$ combinations. However, the practical considerations of system parameters $K, \tau_p,$ and $L$ affect the computational complexity.
\par To solve this optimization, Algorithm 1 initializes $\mathbf{S}$ to zero, having no initial associations. The algorithm comprises three parts: \ac{UE} preference-based association, \ac{AP} preference, and handling repeated associations. For \ac{UE} preference, each \ac{UE} selects the \ac{AP} $l^*$ with the strongest \ac{LSFC}, updating $\mathbf{S}$. \ac{AP} preference involves each \ac{AP} identifying a subset of \ac{UEs} based on \ac{LSFC}, selecting up to $(\tau_p - |\mathcal{M}_l|)$ \ac{UEs}. Handling repeated associations ensures each \ac{UE} is served by at least two \ac{APs} without exceeding the \ac{AP} capacity. The final $\mathbf{S}$ matrix guarantees each \ac{UE} is served by at least two \ac{APs}, without surpassing $\tau_p$, and efficiently resolves the NP-hard problem.
%%%%%%%%%%%%%%%%%%%%%%%%%%%%%%%%%%%%%%%%%%%%%%%%%%%%%%%%%%%%%%%%%%%%%%%%%%%%%%%%%
\begin{algorithm}
\caption{Second Phase \ac{UA}}
\begin{algorithmic}[1]
\STATE \textbf{Input:} $L, K, \{\theta_{i,k}\}, \tau_p, \mathbf{u_k}$, Initial association matrix $\mathbf{S}$ from Algorithm 1
\STATE \textbf{Output:} Refined association matrix $\mathbf{S}$
\FOR{$k = 1:K$}
    \STATE $\mathcal{V}_k = \{ l \ | \ \mathbf{S}(k, l) = 1 \}$ \COMMENT{Initial APs associated with UE $k$}
    \FOR{$i \in \mathcal{V}_k$}
        \STATE Compute $\theta_{i,k}$ \COMMENT{Angle from AP $i$ to UE $k$}
    \ENDFOR
    \STATE Check if angles converge at the estimated UE location
    \IF{angles do not converge}
        \STATE Remove $i$ from $\mathcal{V}_k$
        \STATE Update $\mathbf{S}(k, i) \leftarrow 0$
    \ENDIF
\ENDFOR
\FOR{$k = 1:K$}
    \STATE $\mathcal{V}_k = \{ l \ | \ \mathbf{S}(k, l) = 1 \}$ 
    \IF{$|\mathcal{V}_k| < 2$}
        \STATE $l^* = \arg\min_l \theta_{i,k} \text{ where } l \notin \mathcal{V}_k$
        \STATE $\mathbf{S}(k, l^*) \leftarrow 1$
    \ENDIF
\ENDFOR
\RETURN $\mathbf{S}$
\end{algorithmic}
\end{algorithm}
%%%%%%%%%%%%%%%%%%%%%%%%%%%%%%%%%%%%%%%%%%%%%%%%%%%%%%%%%%%%%%%%%%%%%%%%%%%%%%%%% 
\par To achieve optimal \ac{JRC} \ac{UA}, it is crucial to consider the sensing aspects beyond \ac{SE} of the communication system. For sensing, the echoes must be received solely from the \ac{UE}, avoiding clutter. Thus, the association must ensure that each \ac{UE} is served by \ac{APs} providing the highest \ac{SCNR} while maintaining \ac{LoS} conditions through \ac{AOA} convergence. This convergence is achieved by exploiting \ac{AOA} based positioning \cite{nguyen2013new, 7346436}, which identifies the \ac{APs} that have clutter in their view angle with a \ac{UE}. A heuristic-based solution is proposed here to maximize the minimum \ac{SCNR} while ensuring at least two \ac{APs} serve each \ac{UE} with \ac{LoS} conditions. This approach is executed after forming the initial \ac{LSFC}-based cluster. The cluster is then refined by eliminating \ac{APs} unsuitable for sensing. The formulation is
%%%%%%%%%%%%%%%%%%%%%%%%%%%%%%%%%%%%%%%%%%%%%%%%%%%%%%%%%%%%%%%%%%%%%%%%%%%%%%%%%
\begin{equation}
\label{P2}
\begin{aligned}
\mathbf{P2:} & \max_{\mathbf{S}} \min_{k \in \{1, \ldots, K\}} \left\{ \mathrm{SCNR}_k \right\} \\
\mathrm{s.t.} & \quad |\mathcal{V}_{k}| \geq 2, \quad \forall k = 1, \ldots, K \\
& \quad \theta_{i,k} \text{ ensures LoS conditions, } \forall i \in \mathcal{V}_k \\
& \quad \theta_{i,k} \text{ from APs to UE } \text{ converge at } u_{k}, \forall i \in \mathcal{V}_k 
\end{aligned}
\end{equation}
%%%%%%%%%%%%%%%%%%%%%%%%%%%%%%%%%%%%%%%%%%%%%%%%%%%%%%%%%%%%%%%%%%%%%%%%%%%%%%%%%
\par To solve this optimization problem, the detailed pseudo-code is provided in Algorithm 2. This algorithm begins with the initial association matrix $\mathbf{S}$ from Algorithm 1, which is filled based on the initial association via \ac{LSFC}. In the second step, the angulation-based scheme is applied to the same \ac{APs} that were associated in the first step. This ensures that the association is not restarted but refined, by removing those \ac{APs} that are not suitable for sensing. The method ensures that each \ac{UE} is served by \ac{APs} that provide the highest \ac{SCNR} which is detailed in Section \ref{aoaclutter}. Moreover, both \textbf{P1} and \textbf{P2} are summarized in a flowchart as illustrated in Fig. \ref{flowcc}. 

\begin{figure*}
\centering 
\resizebox{2\columnwidth}{!}{
\includegraphics{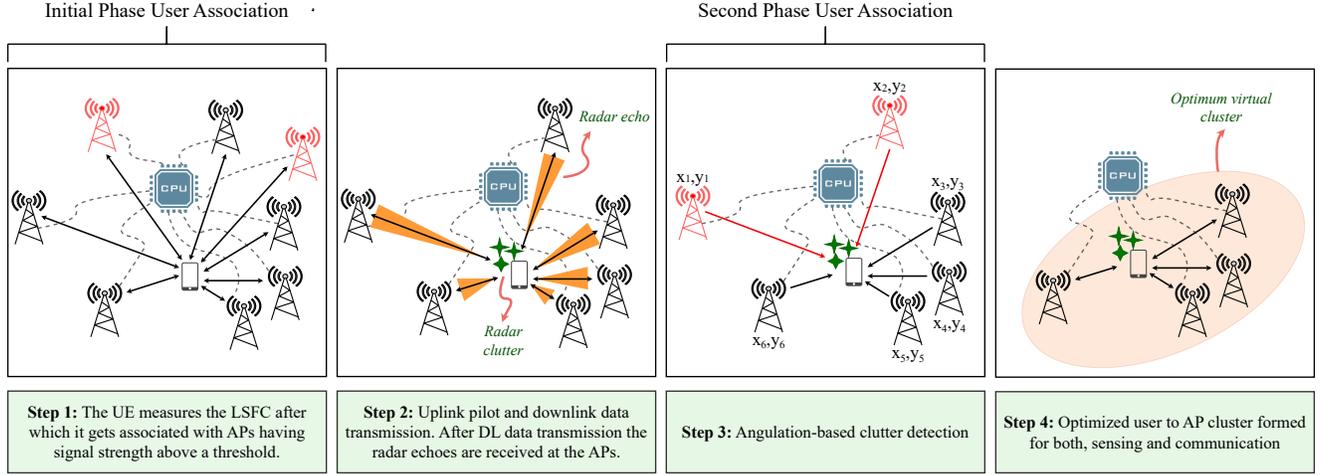}}
\caption{Block diagram of the proposed \ac{UA} technique.}
\label{techhh}
\end{figure*}
%%%%%%%%%%%%%%%%%%%%%%%%%%%%%%%%%%%%%%%%%%%%%%%%%%%%%%%%%%%%%%%%%%%%%%%%%%%%%%%%%
\section{Proposed User Association}
\par Despite the potential benefits of \ac{CF-mMIMO}, effective \ac{UA} is essential to optimize system performance. Traditional \ac{UA} methods may not fully leverage \ac{CF-mMIMO} capabilities in a \ac{JRC} scenario, as they focus solely on communication metrics such as signal strength or \ac{SINR}. This problem now involves dynamically assigning a \ac{UE} to \ac{APs} based on communication and sensing considerations. The key challenge is to develop a scheme that effectively balances communication performance and sensing accuracy. Our proposed \ac{UA} technique, shown in Fig. \ref{techhh}, addresses this issue.
%%%%%%%%%%%%%%%%%%%%%%%%%%%%%%%%%%%%%%%%%%%%%%%%%%%%%%%%%%%%%%%%%%%%%%%%%%%%%%%%%
\vspace{-3mm}
\subsection{Initial Phase for User Association} \label{initial}
\par For each \ac{UE}-\ac{AP} pair, the specific \ac{UE} first measures the $\beta_{kl}$ for all nearby \ac{APs}. This is done using the primary and secondary synchronization signals, a standard feature in cellular networks, and is broadcasted periodically to facilitate such measurements \cite{bjornson2020scalable}. By measuring $\beta_{kl}$, the \ac{UE} selects the highest values and associates with those \ac{APs}. Let $\mathcal{L}_{\text{AP},k}$ denote the set of \ac{AP}s available for the $k$-$th$ \ac{UE}. The \ac{UE} aims to find the $l$-$th$ \ac{AP} that maximizes the value $\beta_{kj}$, given by $l = \arg\max_{j \in \mathcal{L}_{\text{AP},k}} \beta_{kl}$, where the index $j$ ranges over $\mathcal{L}_{\text{AP},k}$, representing the available \ac{AP}s. Moreover, we can rewrite $\mathcal{M}_k$ which is the subset of selected \ac{APs},
$\mathcal{M}_k = \left\{ j \in \mathcal{L}_{\text{AP},k} : \beta_{kj} \geq \xi \right\}$, where $\xi$ is a threshold for minimum acceptable $\beta_{kl}$. Finally, the \ac{UE} associates with the \ac{AP}s in $\mathcal{M}_k$ ensuring the criteria for optimal communication performance.
%%%%%%%%%%%%%%%%%%%%%%%%%%%%%%%%%%%%%%%%%%%%%%%%%%%%%%%%%%%%%%%%%%%%%%%%%%%%%%%%%
\vspace{-5mm}
\subsection{Second Phase for User Association:} \label{aoaclutter}
\par Once the initial phase \ac{APs} are selected, the next step follows by the \ac{UL} pilots transmission and \ac{APs} transmitting \ac{DFRC} beam. When the \ac{DFRC} beam is directed towards the \ac{UE}, reflected radar echoes are received at each \ac{AP}. These echoes help estimate the \ac{UEs} parameters such as its range, velocity, and angular estimates, as described in \cite{10458884}. In this work, the \ac{AOA} estimates from the \ac{DFRC} echo received by the \ac{APs} are exploited to detect clutter around the desired target, resulting in the exclusion of \ac{APs} with clutter in their view angles.
\par
\begin{figure}
\centering 
\resizebox{1\columnwidth}{!}{
\includegraphics{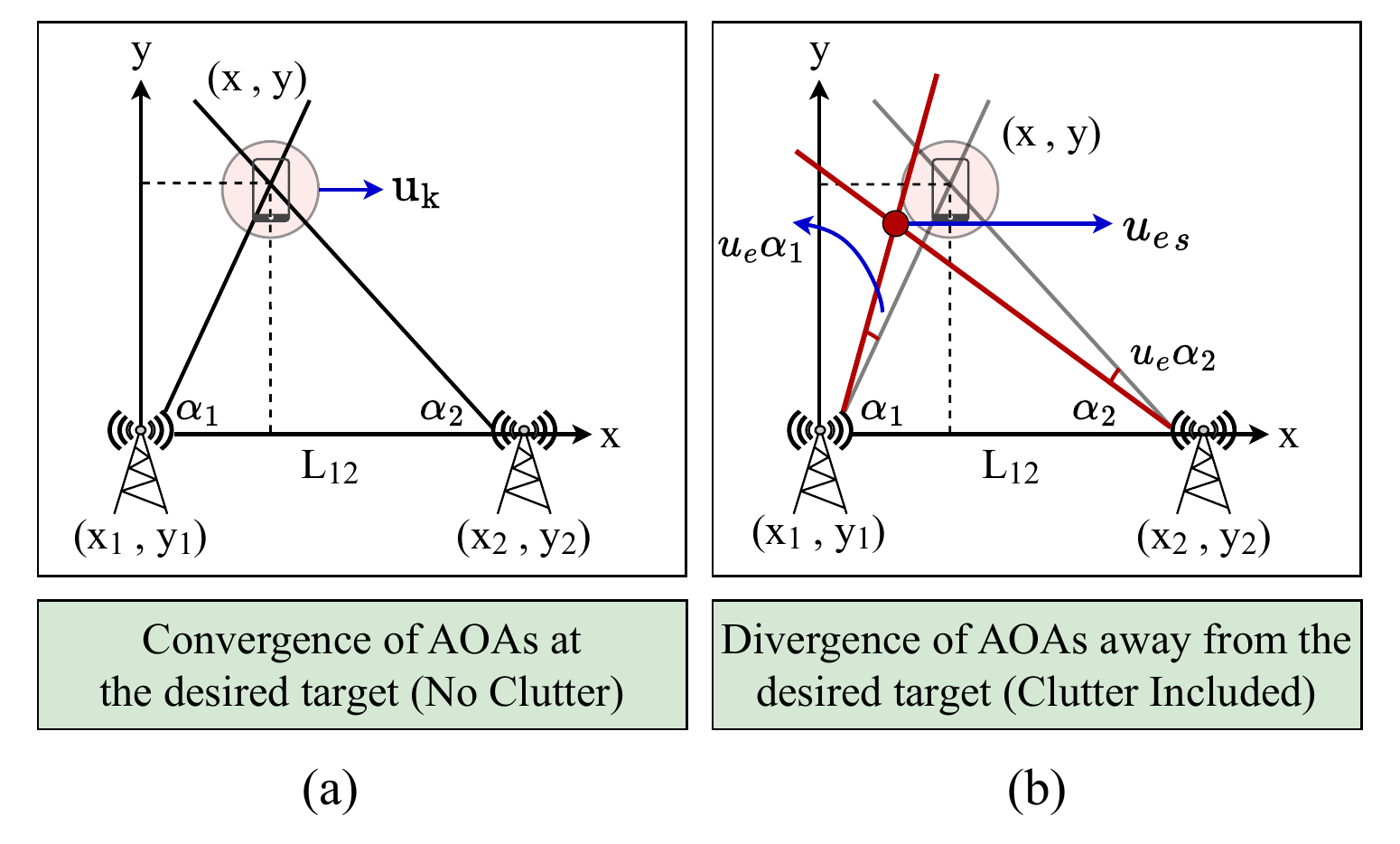}}
\caption{Angulation-based clutter detection technique.}
\label{aoa}
\end{figure}

 In Cartesian coordinate system let $\mathbf{u_k^{\prime}}=[x~y]^T$ be the real position of the \ac{UE} and $\mathbf{p}_l=[x_l~y_l]^T$, $(l$ = $1,\cdots,n)$ the position of the $l$-$th$ \ac{AP}. Without loss of generality, the position of one of the \ac{APs} is considered at origin i.e., $\mathbf{p}_1=[x_1~y_1]^T\equiv[0~0]^T$. Moreover, $\alpha_i$ is the \ac{AOA} estimated from echo at $l$-$th$ \ac{AP}. Let $L_{1d}$ be the distance between \ac{AP}$_1$ and \ac{AP}$_d$, $(d=2,\cdots,n)$, as illustrated in Fig. \ref{aoa}. Since the network already knows the locations of each \ac{AP}, using the abscissa and the ordinate \cite{fish1909coordinates}, \cite{stewart2011calculus} and combining the estimated \ac{AOA}s, the position where the \ac{AOA} bearings intersect at $\mathbf{u_k^{\prime}}$ is $\begin{bmatrix}x\\y\end{bmatrix}=\begin{bmatrix}\dfrac{\cos\alpha_1\sin\alpha_n}{\sin(\alpha_1+\alpha_n)}L_{1d}\\\dfrac{\sin\alpha_1\sin\alpha_n}{\sin(\alpha_1+\alpha_n)}L_{1d}\end{bmatrix}$, \cite{nguyen2013new}. Combining all these equations leads to $\mathbf{G}\mathbf{u_k^{\prime}}=\mathbf{b}$. Since it is linearized, the least square estimator provides the intersection estimate, $\mathbf{u_k^\prime}_{(\textrm{LS})}=(\mathbf{G}^T\mathbf{G})^{-1}\mathbf{G}^T\mathbf{b}$, where
%%%%%%%%%%%%%%%%%%%%%%%%%%%%%%%%%%%%%%%%%%%%%%%%%%%%%%%%%%%%%%%%%%%%%%%%%%%%%%%%% 

{\small
\begin{equation}
\begin{aligned}
&\mathbf{u_k^{\prime}}=\begin{bmatrix}x\\y\end{bmatrix},
&\mathbf{G}=
\begin{bmatrix}1&0\\0&1\\1&0\\0&1\\\vdots&\vdots\\1&0\\0&1
\end{bmatrix},
\mathbf{b}=\begin{bmatrix}\frac{\cos(\alpha_1)\sin(\alpha_2)}{\sin(\alpha_1+\alpha_2)}L_{12}\\\frac{\sin(\alpha_1)\sin(\alpha_2)}{\sin(\alpha_1+\alpha_2)}L_{12}\\\vdots\\\frac{\cos(\alpha_1)\sin(\alpha_n)}{\sin(\alpha_1+\alpha_n)}L_{1d}\\\frac{\sin(\alpha_1)\sin(\alpha_n)}{\sin(\alpha_1+\alpha_n)}L_{1d}\end{bmatrix}.
\end{aligned}
\end{equation}}

\par In case when there is clutter in front of a specific \ac{AP} view angle with the \ac{UE}, the \ac{AOA} will be corrupted and will have wrong estimates. Therefore the error measurements will be ${u_e}\alpha_1$ and ${u_e}\alpha_2$ instead of $\alpha_1$ and $\alpha_2$ respectively as shown in Fig. \ref{aoa} (b). Moreover, the term ${u_e}_s$ in the abscissa and the ordinate is expressed as ${u_e}_s=dxdy=\left|\frac{\partial(x,y)}{\partial(\alpha_1,\alpha_2)}\right|{u_e}\alpha_1{u_e}\alpha_2$. Thus, the derivatives of $x$ and $y$ with respect to their \ac{AOA}s are $\begin{bmatrix}\dfrac{\partial x}{\partial\alpha_1}&\dfrac{\partial x}{\partial\alpha_2}\\\dfrac{\partial y}{\partial\alpha_1}&\dfrac{\partial y}{\partial\alpha_2}\end{bmatrix}=\begin{bmatrix}\dfrac{-\sin\alpha_2\cos\alpha_2}{\sin^2(\alpha_1+\alpha_2)}&\dfrac{\sin\alpha_1\cos\alpha_1}{\sin^2(\alpha_1+\alpha_2)}\\\dfrac{\sin^2\alpha_2}{\sin^2(\alpha_1+\alpha_2)}&\dfrac{\sin^2\alpha_1}{\sin^2(\alpha_1+\alpha_2)}\end{bmatrix}L_{12}$, $\left|\dfrac{\partial(x,y)}{\partial(\alpha_1,\alpha_2)}\right|=\left|\dfrac{\sin\alpha_1\sin\alpha_2}{\sin^3(\alpha_1+\alpha_2)}\right|L_{12}^2$, replacing this into the ${u_e}_s$ equation yields ${u_e}_s=L_{12}^{2}\left|\frac{\sin\alpha_{1}\sin\alpha_{2}}{\sin^{3}(\alpha_{1}+\alpha_{2})}\right|d\alpha_{1}d\alpha_{2}.$
In this context, the point ${u_e}_s$ represents the intersection of the \ac{AOA}s from two \ac{AP}s, which were inaccurately estimated due to clutter obstructing the view between the \ac{AP} and the \ac{UE}. The requirement of at least two \ac{AP}s connected to a \ac{UE} allows for the intersection of their angles. If the angles intersect around the coarse position $\mathbf{u_k^{\prime}}$ of the \ac{UE}, it indicates no clutter in front of this specific \ac{AP}. Suppose they intersect elsewhere, as illustrated in Fig. \ref{aoa} (b), in that case, it suggests that the \ac{AOA} estimate from the radar echo was affected by clutter, causing the incorrect \ac{AOA} estimation to deviate from $\mathbf{u_k^{\prime}}$. This method allows us to identify any \ac{AP} obstructed by clutter, enabling its removal from the \ac{AP} cluster. By doing so, the echoes received are solely from the target, eliminating interference from surrounding clutter. This approach increases \ac{LoS} links, thereby enhancing the \ac{SCNR} and overall sensing performance.

%%%%%%%%%%%%%%%%%%%%%%%%%%%%%%%%%%%%%%%%%%%%%%%%%%%%%%%%%%%%%%%%%%%%%%%%%%%%%%%%%
%%%%%%%%%%%%%%%%%%%%%%%%%%%%%%%%%%%%%%%%%%%%%%%%%%%%%%%%%%%%%%%%%%%%%%%%%%%%%%%%%
%%%%%%%%%%%%%%%%%%%%%%%%%%%%%%%%%%%%%%%%%%%%%%%%%%%%%%%%%%%%%%%%%%%%%%%%%%%%%%%%%
% \begin{figure}[hbt!] 
% \centering 
% \resizebox{1\columnwidth}{!}{
% \includegraphics{Figures/trylast.eps}}
% \caption{2D distribution of the locations of the \ac{APs}, \ac{UEs} and optimal \ac{AP} cluster (green circles represent the final optimized \ac{APs} while the \ac{APs} which have clutter in front of their view angle are represented by the red cross, which are excluded from the final cluster in the second step).}
% \label{2dd}
% \end{figure}
%%%%%%%%%%%%%%%%%%%%%%%%%%%%%%%%%%%%%%%%%%%%%%%%%%%%%%%%%%%%%%%%%%%%%%%%%%%%%%%%%
%%%%%%%%%%%%%%%%%%%%%%%%%%%%%%%%%%%%%%%%%%%%%%%%%%%%%%%%%%%%%%%%%%%%%%%%%%%%%%%%%
\begin{figure*}[ht]
\centering      
\subfloat[]{
  \includegraphics[width=58mm]{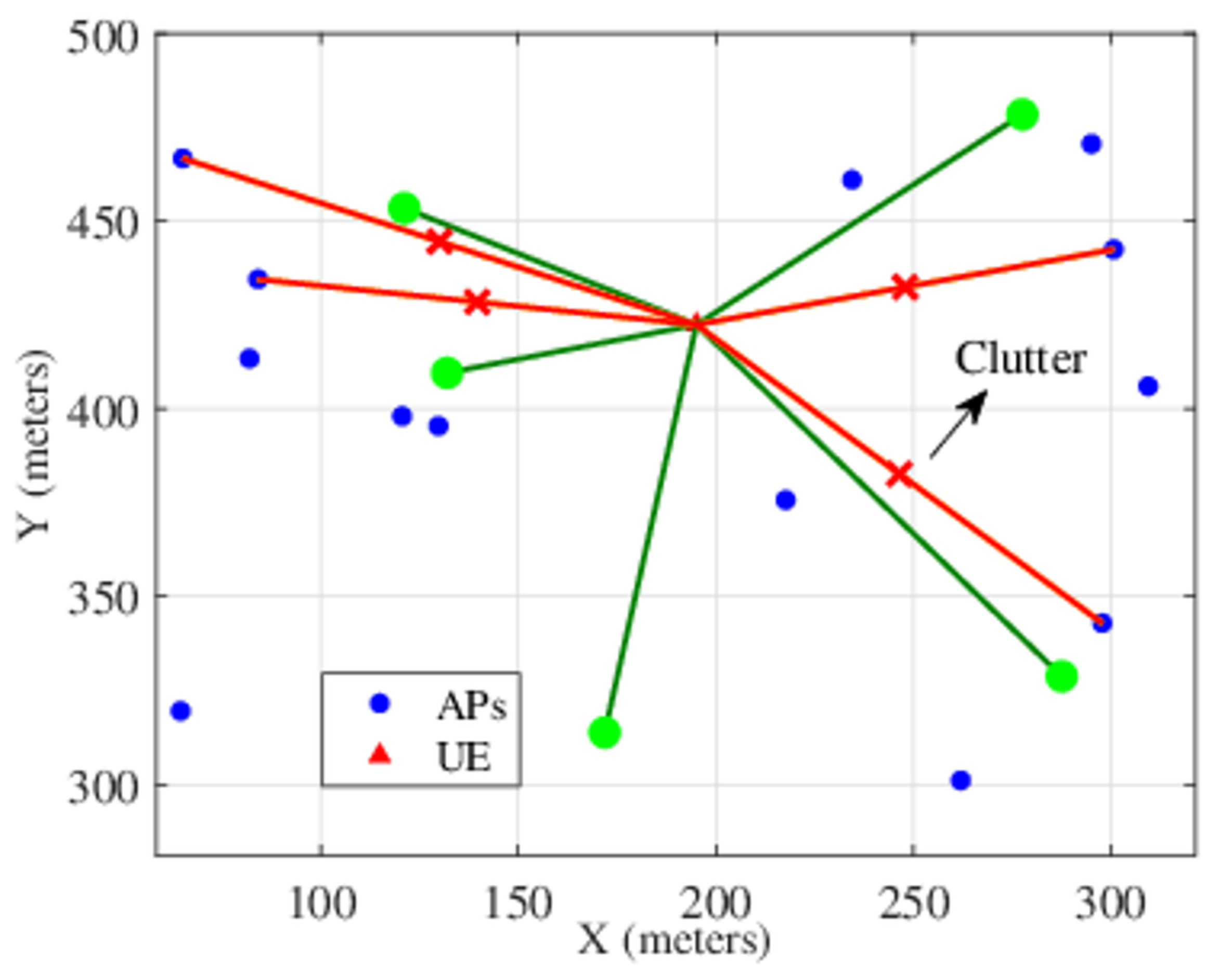}
}
\subfloat[]{
  \includegraphics[width=58mm]{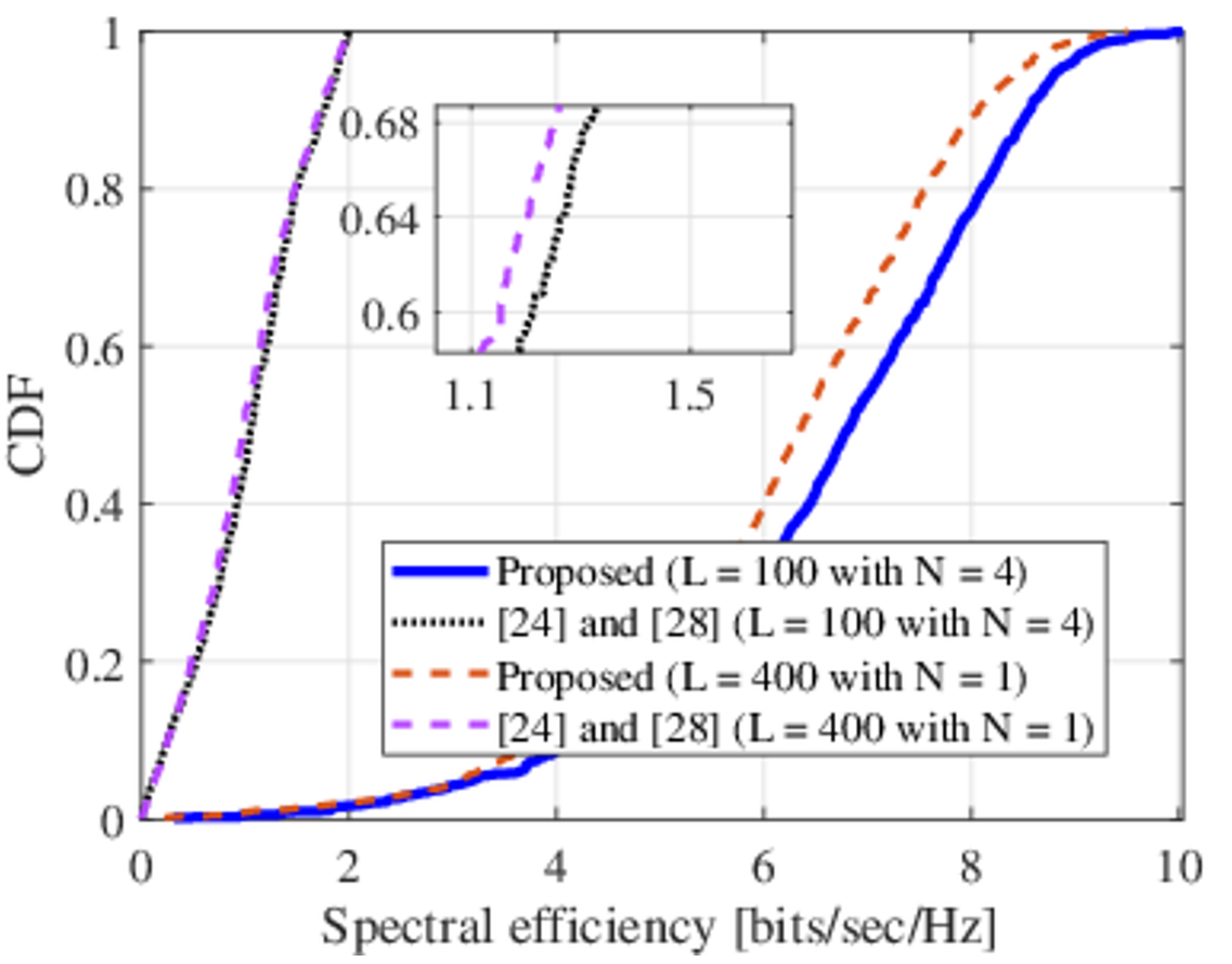}
}
\subfloat[]{
  \includegraphics[width=58mm]{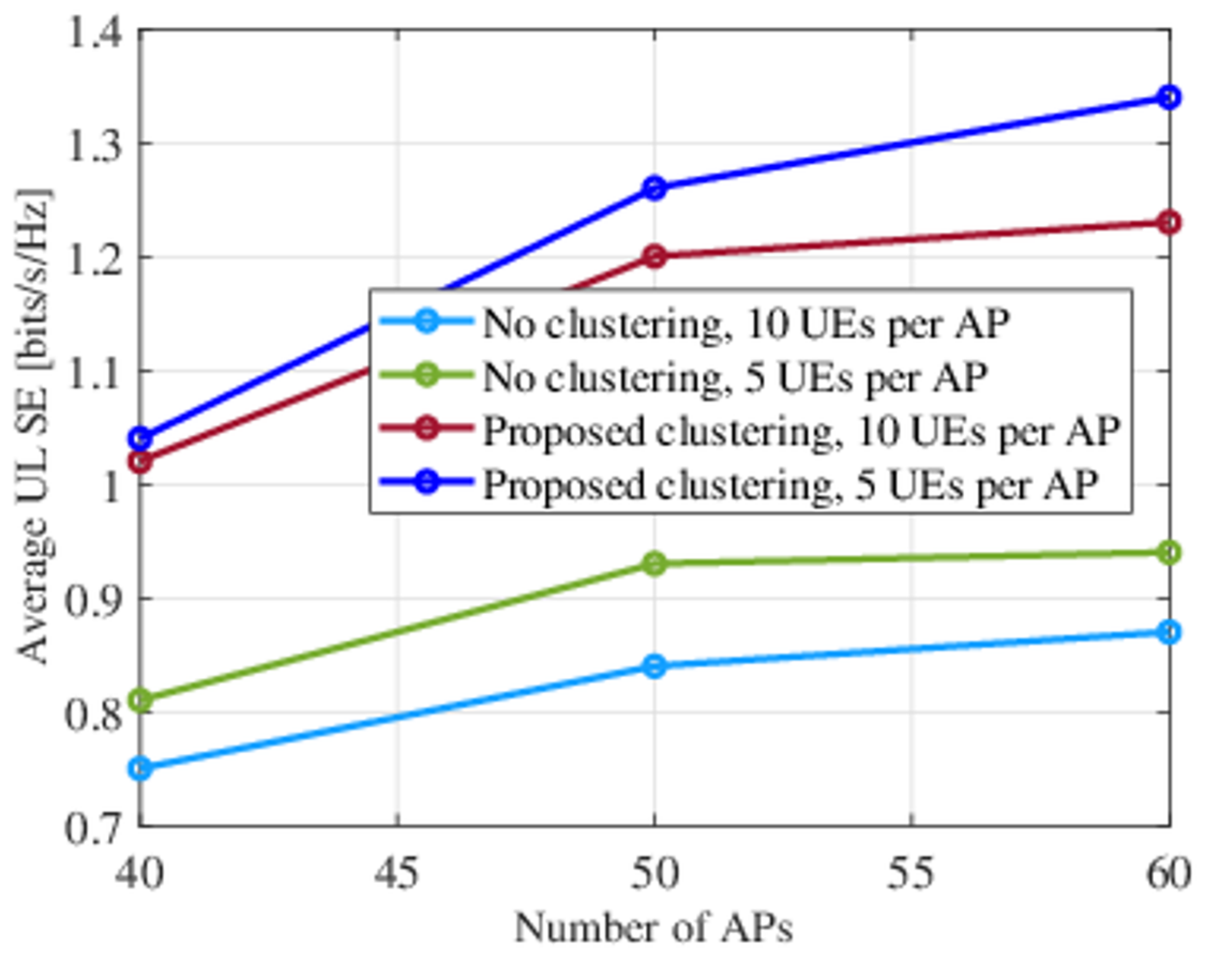}
}

\hspace{0mm}
\subfloat[]{
  \includegraphics[width=58mm]{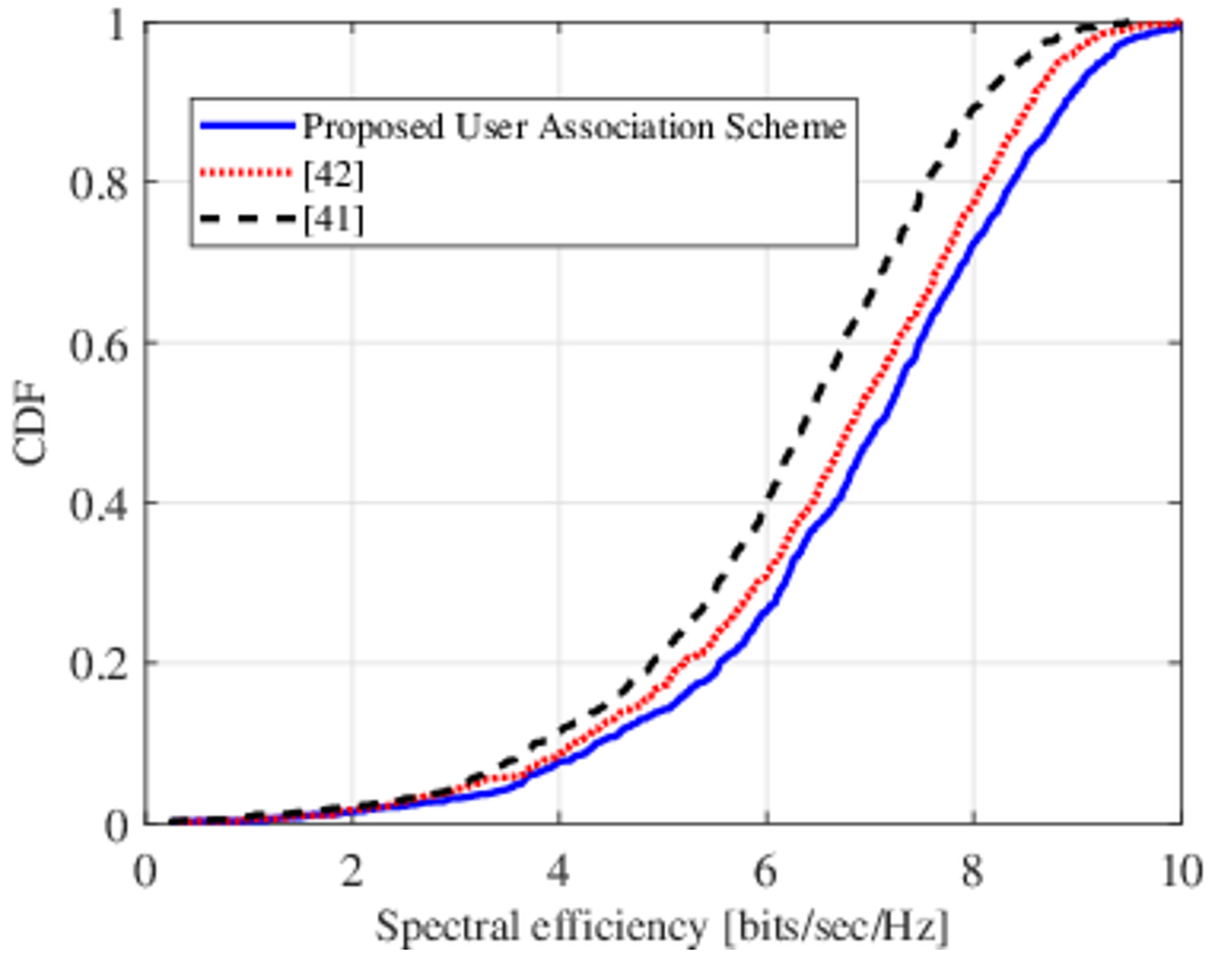}
}
\subfloat[]{
  \includegraphics[width=58mm]{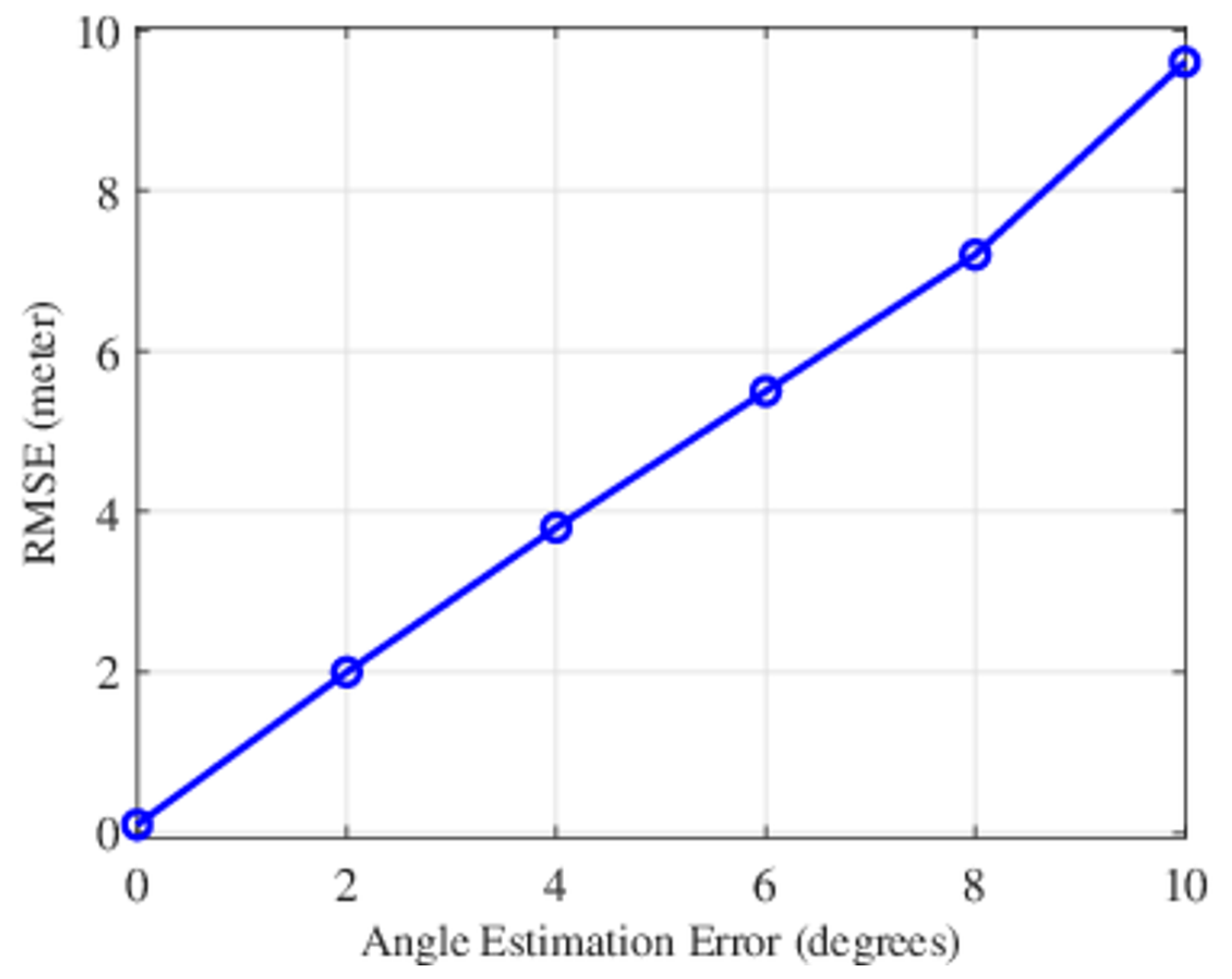}
}
\subfloat[]{
  \includegraphics[width=58mm]{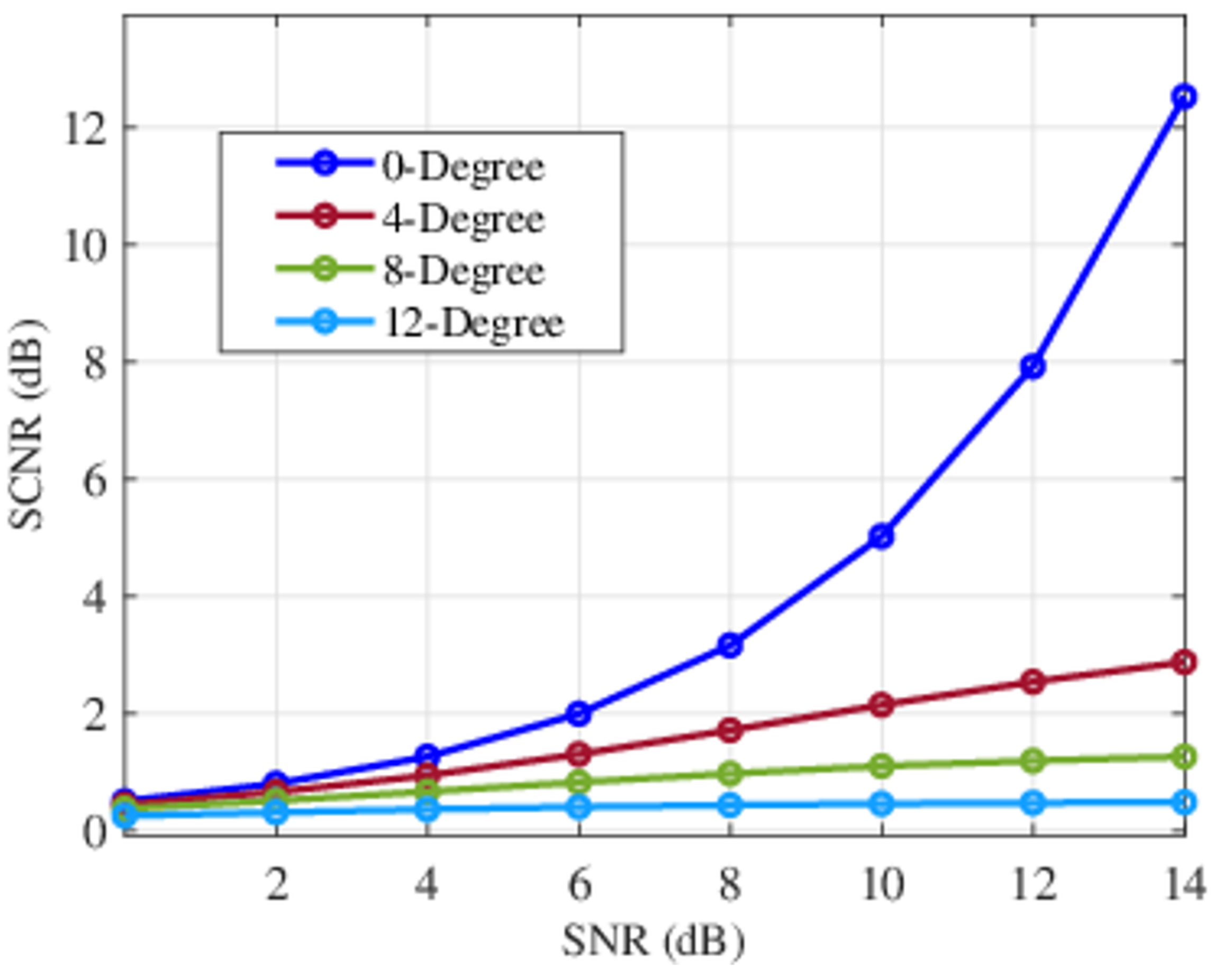}
}
\caption{(a) 2D distribution of the \ac{APs}, \ac{UEs} and optimal \ac{AP} cluster (green circles are the final optimized \ac{APs} while the \ac{APs} which have clutter in front of their view angle are represented by the red cross, which are excluded from the final cluster), (b) \ac{SE} comparison per \ac{UE} for scalable and non-scalable \ac{CF-mMIMO}, (c) illustrates the effect of number of \ac{APs} vs average \ac{UL} \ac{SE} for proposed clustering method and for scenario where no clustering is performed, (d) \ac{CDF} of \ac{SE} of association schemes, (e) \ac{RMSE} vs. angle estimation error due to clutter and (f) illustrates \ac{SCNR} vs \ac{SNR} for different error in \ac{AOA} estimation.}
\label{MIANthing22}
\end{figure*}
%%%%%%%%%%%%%%%%%%%%%%%%%%%%%%%%%%%%%%%%%%%%%%%%%%%%%%%%%%%%%%%%%%%%%%%%%%%%%%%%%
\vspace{-3mm}
\subsection{Final Optimized Cluster}
\par Once the \ac{APs} designated to serve a specific \ac{UE} are determined by following the steps outlined in Section \ref{initial} and \ref{aoaclutter}, as illustrated in Fig. \ref{techhh}, these selected \ac{APs} will continuously sense and communicate with the \ac{UE}. For the communication process, conventional \ac{CF-mMIMO} transmission, reception, and decoding techniques are employed, as detailed in \cite{bjornson2020scalable}, \cite{chen2020structured}, and \cite{10310167}, which are also discussed in Section \ref{sysModel}. Meanwhile, we use the following optimal sensing detector to track the \ac{UE} constantly. The reflected echo signals from the \ac{UE} to all the \ac{APs} are sent to the \ac{CPU} for combined detection. Using all these received signals, the detection leads to the following binary hypothesis problem \cite{6601713}

%%%%%%%%%%%%%%%%%%%%%%%%%%%%%%%%%%%%%%%%%%%%%%%%%%%%%%%%%%%%%%%%%%%%%%%%%%%%%%%%%
\begin{equation}
\left\{\begin{array}{ll}
\mathrm{H}_0: & \mathbf{y} = \mathbf{c} + \mathbf{n} \\
\mathrm{H}_1: & \mathbf{y} = \mathbf{x} + \mathbf{c} + \mathbf{n}
\end{array}\right.
\end{equation}
%%%%%%%%%%%%%%%%%%%%%%%%%%%%%%%%%%%%%%%%%%%%%%%%%%%%%%%%%%%%%%%%%%%%%%%%%%%%%%%%%
where $\mathbf{y}, \mathbf{x}, \mathbf{c},$ and $\mathbf{n}$ are defined by column-wise stacking of $\mathbf{y}_l, \mathbf{x}_l, \mathbf{c}_l,$ and $\mathbf{n}_l$ for $l = 1, 2, \ldots, N$. More precisely, $\mathbf{y} \triangleq [\mathbf{y}_1^T \cdots \mathbf{y}_N^T]^T$, $\mathbf{x} \triangleq [\mathbf{x}_1^T \cdots \mathbf{x}_N^T]^T$, $\mathbf{c} \triangleq [\mathbf{c}_1^T \cdots \mathbf{c}_N^T]^T$, and $\mathbf{n} \triangleq [\mathbf{n}_1^T \cdots \mathbf{n}_N^T]^T$. Let $\{\mathbf{N}_l\}$ denote the covariance matrices of Gaussian random vectors $\{\mathbf{n}_l\}$. Further let $\mathbf{X}, \mathbf{C},$ and $\mathbf{N}$ represent the covariance matrices of $\mathbf{x}, \mathbf{c},$ and $\mathbf{n}$ respectively. Using the assumptions mentioned earlier we have $\mathbf{X} = \mathrm{blkDiag}(\sigma_1^2 \mathbf{a} \mathbf{a}^H, \sigma_2^2 \mathbf{a} \mathbf{a}^H, \ldots, \sigma_N^2 \mathbf{a} \mathbf{a}^H)$, $\mathbf{C} = \mathrm{blkDiag}(\sigma_{c,1}^2 \mathbf{a} \mathbf{a}^H, \sigma_{c,2}^2 \mathbf{a} \mathbf{a}^H, \ldots, \sigma_{c,N}^2 \mathbf{a} \mathbf{a}^H)$, and $\mathbf{N} = \mathrm{blkDiag}(\mathbf{N}_1, \mathbf{N}_2, \ldots, \mathbf{N}_N)$. Consequently, we have
%%%%%%%%%%%%%%%%%%%%%%%%%%%%%%%%%%%%%%%%%%%%%%%%%%%%%%%%%%%%%%%%%%%%%%%%%%%%%%%%%
\begin{equation} \label{detector}
\left\{\begin{matrix}
\mathrm{H}_0: & \mathbf{x} \sim \mathcal{CN}(\mathbf{0}, \mathbf{I}) \\
\mathrm{H}_1: & \mathbf{x} \sim \mathcal{CN}(\mathbf{0}, \mathbf{DSD} + \mathbf{I})
\end{matrix}\right.
\end{equation}
%%%%%%%%%%%%%%%%%%%%%%%%%%%%%%%%%%%%%%%%%%%%%%%%%%%%%%%%%%%%%%%%%%%%%%%%%%%%%%%%%
where $\mathbf{D} \triangleq (\mathbf{C} + \mathbf{N})^{-\frac{1}{2}} = \mathrm{blkDiag}(\mathbf{D}_1, \mathbf{D}_2, \ldots, \mathbf{D}_N)$ with $\mathbf{D}_k \triangleq (\sigma_{c,k}^2 \mathbf{a} \mathbf{a}^H + \mathbf{N}_k)^{-\frac{1}{2}}$ and $\mathbf{x} \triangleq \mathbf{Dr}$. Note that $\mathbf{D}$ and $\mathbf{X}$ in (\ref{detector}) depend on $\mathbf{a}$. The optimal detector for (\ref{detector}) can be obtained by applying the estimator-correlator theorem as $\sum_{k=1}^N \sigma_k^2 \mathbf{x}_k^H \mathbf{D}_k \mathbf{a} \mathbf{a}^H \mathbf{D}_k (\sigma_k^2 \mathbf{D}_k \mathbf{a} \mathbf{a}^H \mathbf{D}_k + \mathbf{I})^{-1} \mathbf{x}_k \overset{\mathrm{H}_0}{\operatorname*{\leqslant}} \eta$ \cite{kay1998statistical}, where $\eta$ is the threshold, and $\mathbf{x}_k = \mathbf{D}_k \mathbf{r}_k$. Further, $\lambda_k \triangleq \sigma_k^2 \mathbf{a}^H \mathbf{D}_k^2 \mathbf{a}$ and $\theta_k \triangleq \frac{\mathbf{a}^H \mathbf{D}_k \mathbf{x}_k}{\|\mathbf{a}^H \mathbf{D}_k\|_2}$, the canonical form of the detector is $T(\boldsymbol{\theta}) \triangleq \sum_{k=1}^N \frac{\lambda_k |\theta_k|^2}{1 + \lambda_k} \overset{\mathrm{H}_0}{\operatorname*{\underset{\mathrm{H}_1}{\operatorname*{\leqslant}}}} \eta$ where $ \boldsymbol{\theta} \triangleq [\theta_1 \theta_2 \cdots \theta_N]^T.$
%%%%%%%%%%%%%%%%%%%%%%%%%%%%%%%%%%%%%%%%%%%%%%%%%%%%%%%%%%%%%%%%%%%%%%%%%%%%%%%%%
\section{Simulation Results} 
\par In this section, simulation results are provided considering a service area of $500\times500$m with $L=100$ and $K$ \ac{UEs}, which are uniformly and independently distributed throughout the region. Every \ac{AP} comprises $N=4$ uniform linear array antennas, with a half-wavelength spacing between any two successive antennas. The 2D locations of the \ac{APs}, \ac{UE}, and the optimal cluster of \ac{APs} are illustrated in Fig. \ref{MIANthing22} (a). The large-scale fading gain is determined by assuming $\Upsilon=-148.1$ dB, $\alpha=3.76, d_{\mathrm{ref}}=1$ km, and $\sigma_{dB}=$ 10 dB \cite{bjornson2017massive}. Furthermore, we adopt system configuration as in \cite{bjornson2020scalable}, bandwidth is $B=20$ MHz and each \ac{UE} can transmit a maximum of 100 mW power. The path loss for the communication channels is modeled using the 3GPP Urban Microcell model, as defined in \cite{3GPP2017} and for the sensing channels is modeled using the two-way radar equation. The sensing detection threshold i.e., $\eta$ is determined according to the false alarm probability of 0.1, which is relevant for radar applications \cite{guruacharya2020map}.
%%%%%%%%%%%%%%%%%%%%%%%%%%%%%%%%%%%%%%%%%%%%%%%%%%%%%%%%%%%%%%%%%%%%%%%%%%%%%%%%%
\vspace{-3mm}
\subsection{Scalability and need for User Association}
\par To highlight the scalability of \ac{CF-mMIMO}, it is essential to focus on \ac{UA} and optimal cluster formation. Early concepts of \ac{CF-mMIMO} in \cite{ngo2017cell} and \cite{7869024} faced scalability issues, assuming all \ac{APs} serving all \ac{UEs} without limiting the number of \ac{UEs} each \ac{AP} could handle. As $K$ grows, this approach becomes impractical, failing to ensure service to all \ac{UEs}. Each \ac{AP} would need to compute channel estimates for all $K$ \ac{UEs}, with infinite complexity as 
$K\to \infty$. Moreover, the signal transmission process would imply infinite complexity. 
\par This section will compare the scalability aspect to show the importance of \ac{UA} and clustering. Fig. \ref{MIANthing22} (b) illustrates the \ac{CDF} vs the \ac{SE} per \ac{UE} when a) $L$ = 400, $N$ = 1 and b) $L$ = 100, $N$ = 4 in case of both the proposed scalable \ac{UA} and non-scalable \ac{CF-mMIMO} as in \cite{ngo2017cell} and \cite{7869024}. We compare the proposed scalable distributed \ac{MMSE} with the non-scalable schemes where all the \ac{APs} serve all the \ac{UEs}. The proposed distributed \ac{MMSE} based \ac{UA} performs well with an average \ac{SE} almost 2.78$\times$ higher than the non-scalable \ac{CF-mMIMO} as can be seen in Fig. \ref{MIANthing22} (b). The main idea is that the nearest \ac{APs} capture the majority of the total received power for a specific \ac{UE}, and these are the \ac{APs} chosen by our clustering algorithm to serve that \ac{UE}. Therefore, it is enough to mitigate any interference between the \ac{UEs} that these \ac{APs} jointly serve. In non-scalable \ac{CF-mMIMO} systems, where all \ac{APs} serve all \ac{UEs}, significant interference arises. However, this issue is mitigated in scalable systems with \ac{UA} and clustering.
\par Moreover Fig. \ref{MIANthing22} (c) illustrates the \ac{UL} \ac{SE} (\ref{bottom}), of the system when \ac{UA} clustering is performed vs when no clustering \ac{UA} is done, that means non-scalable \ac{CF-mMIMO}. Without any clustering, increasing the number of \ac{APs} consistently improves the average user \ac{SE}, regardless of the number of \ac{UEs} per \ac{AP}. However, the proposed clustering schemes, which consider \ac{UA} clustering, demonstrate significant advantages over the no-clustering approach. As the number of \ac{APs} increases, the performance gap between the proposed clustering schemes and the no-clustering scenarios becomes more pronounced, highlighting the benefits of the clustering approach in enhancing \ac{SE}. These findings underscore the importance of opting for \ac{UA} clustering.
%%%%%%%%%%%%%%%%%%%%%%%%%%%%%%%%%%%%%%%%%%%%%%%%%%%%%%%%%%%%%%%%%%%%%%%%%%%%%%%%%
\begin{figure*}
\centering    
\subfloat[]{
  \includegraphics[width=60mm]{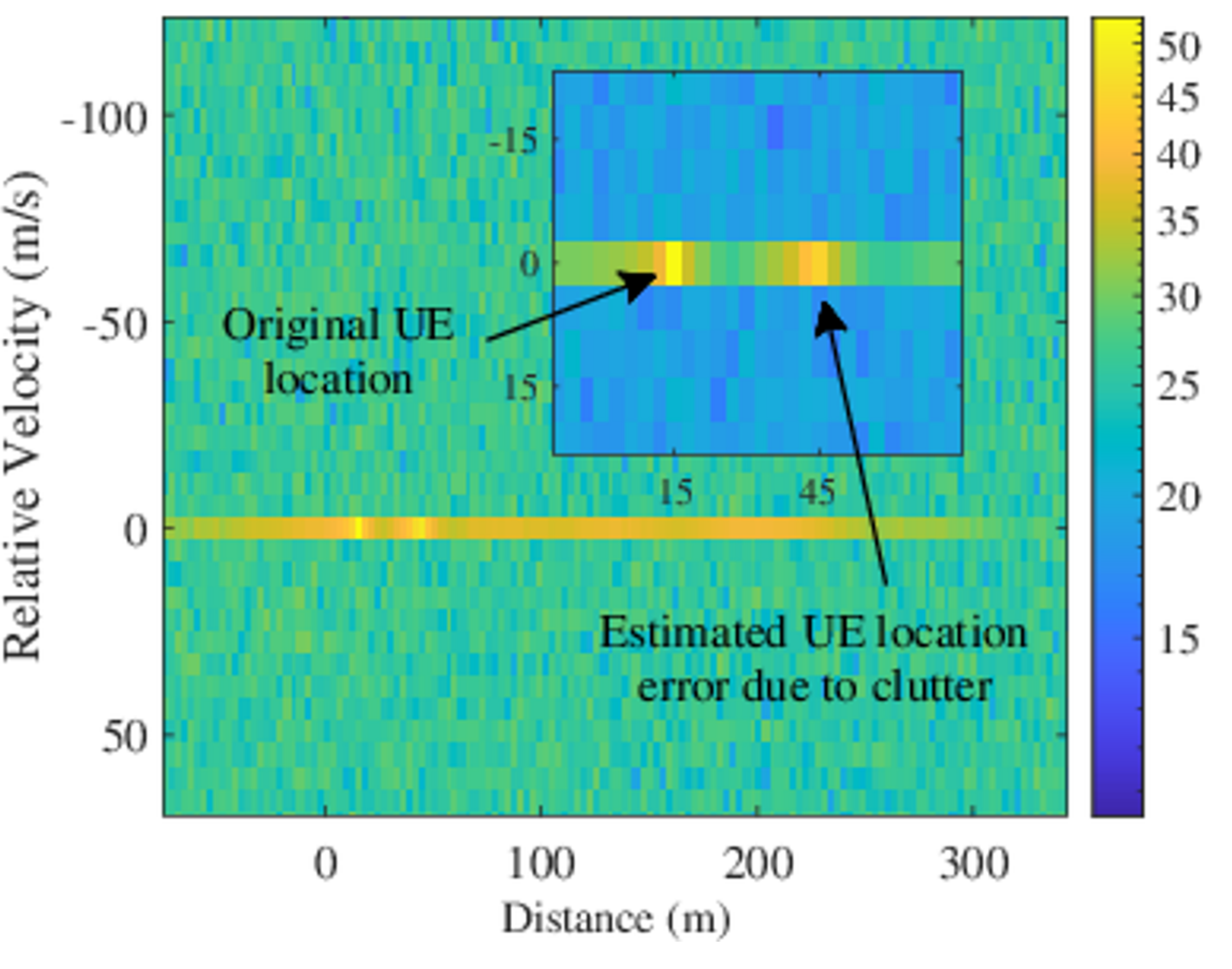}
}
\subfloat[]{
  \includegraphics[width=60mm]{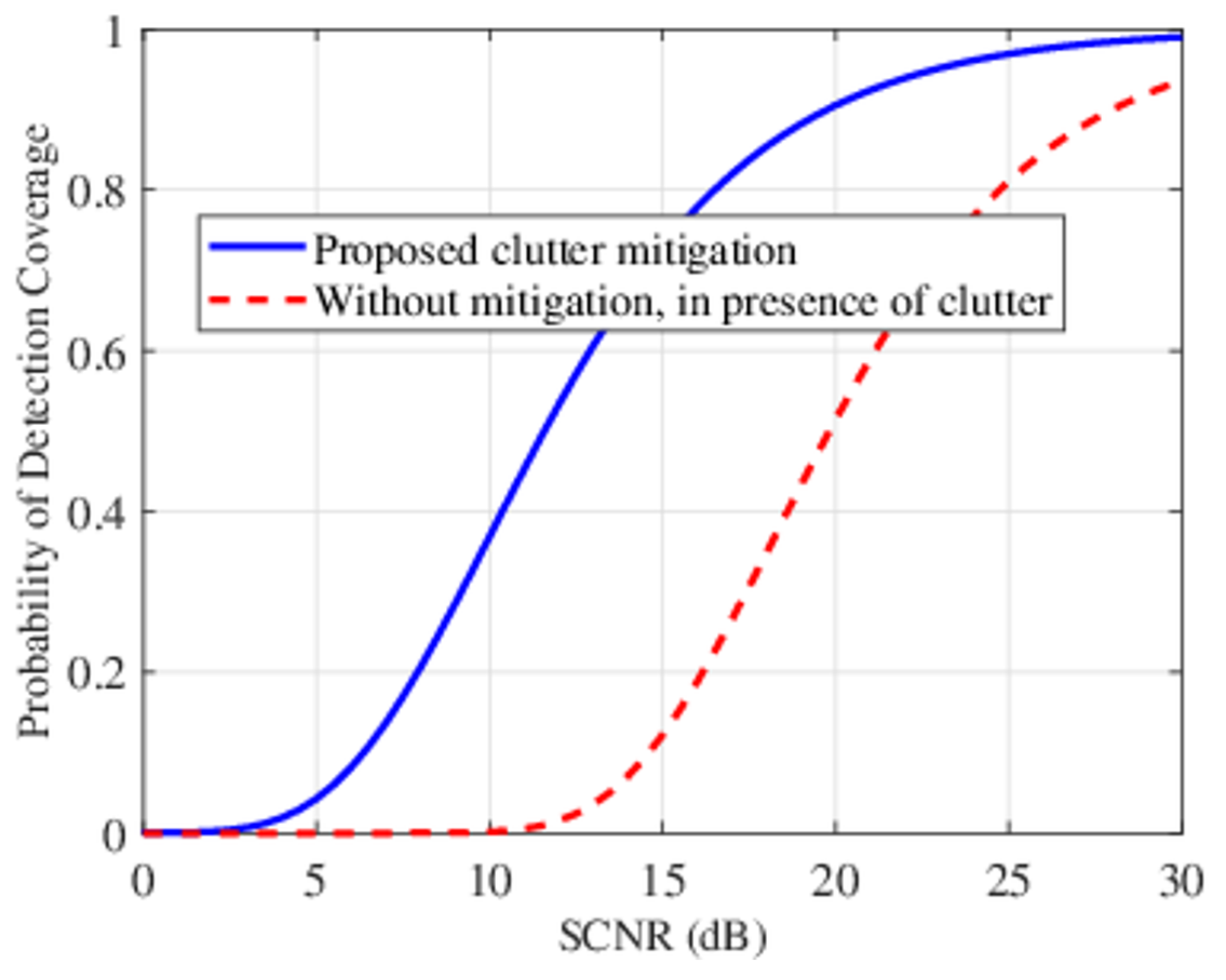}
}
\hspace{0mm}
\subfloat[]{   
  \includegraphics[width=60mm]{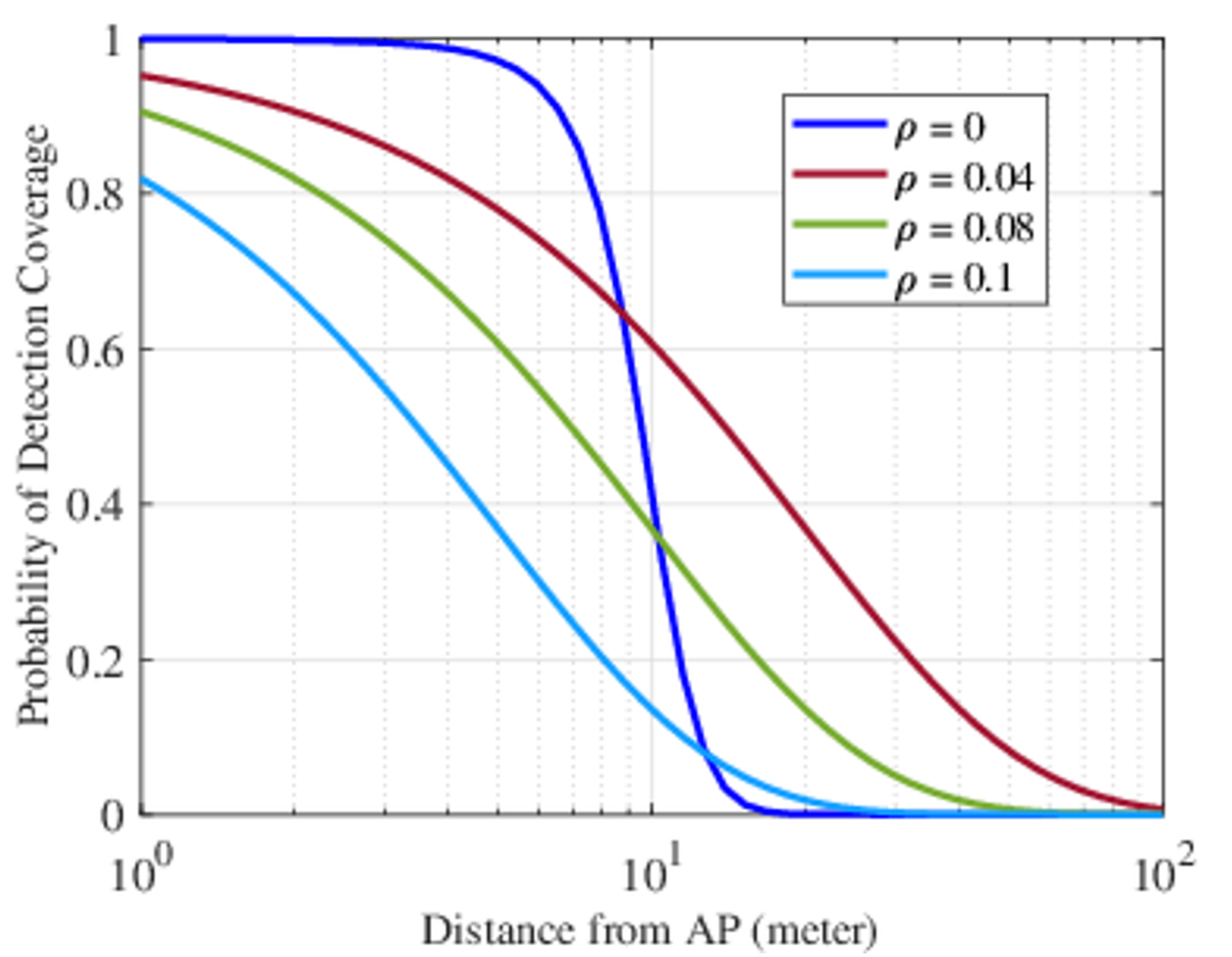}
}
\subfloat[]{
  \includegraphics[width=60mm]{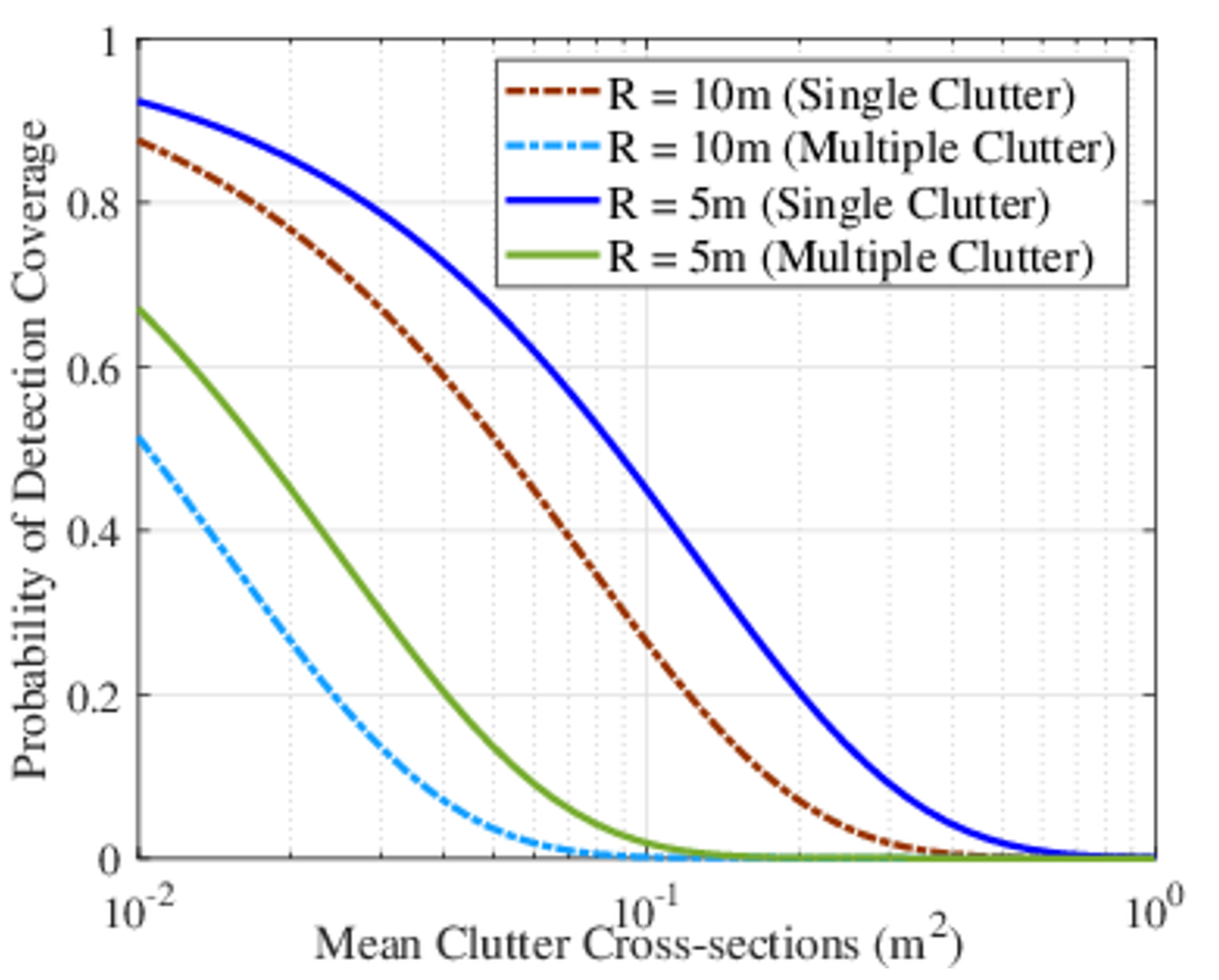}
}
\caption{(a) velocity-distance plot for the original \ac{UE} location and wrongly estimated target location due to the presence of clutter, (b) Probability of detection coverage vs \ac{SCNR}, (c) variation of \ac{$P_{dc}$} with the \ac{UE} distance for various clutter densities, keeping the mean clutter cross-section and radar parameters constant and (d) is the variation of \ac{$P_{dc}$} with \ac{UE} distance for different mean clutter cross-sections for single or multiple clutters.}
\label{MIANTHING}
\end{figure*}
%%%%%%%%%%%%%%%%%%%%%%%%%%%%%%%%%%%%%%%%%%%%%%%%%%%%%%%%%%%%%%%%%%%%%%%%%%%%%%%%
\vspace{-3mm}
\subsection{Performance of the Initial Phase User Association} 
\par In this section, we compare the performance of the initial phase \ac{UA}, with existing schemes from the literature. Specifically, we evaluate our \ac{LSFC} scheme against those presented in \cite{9174860} and \cite{9449014}. The \ac{CDF} of the \ac{SE} per \ac{UE} is illustrated in Fig. \ref{MIANthing22} (d), assuming that $\tau_p$ = 10, $K$ = 50. The scheme in \cite{9174860} exhibits the lowest \ac{SE} per \ac{UE} among the compared methods. This is primarily because the algorithm may have difficulty identifying \ac{APs} that can effectively serve multiple \ac{UEs} with weak channel conditions. Conversely, in \cite{9449014} associates \ac{UEs} with weak channel strengths to \ac{APs} that provide the highest sum of \ac{LSFC} for the associated \ac{UEs} across all \ac{APs}. However, this strategy may not always result in optimal \ac{SE} performance. Our \ac{LSFC} scheme outperforms both of these methods, delivering superior \ac{SE} per \ac{UE}. This demonstrates the effectiveness of our approach in identifying and leveraging \ac{APs} with strong channel conditions. 
%%%%%%%%%%%%%%%%%%%%%%%%%%%%%%%%%%%%%%%%%%%%%%%%%%%%%%%%%%%%%%%%%%%%%%%%%%%%%%%%%
\vspace{-3mm}
\subsection{Performance of the Second Phase User Association}
\par In this section to show the efficiency of the angulation-based clutter mitigation we measure the performance by \ac{RMSE}, defined as $\mathrm{RMSE}=\sqrt{\frac{1}{Q}\sum_{i=1}^{Q}||\mathrm{UE}_{\mathrm{estimate}}^{i}-\mathrm{UE}_{\mathrm{real}}||^{2}},$ where $Q$ is the number of measurements for the \ac{UEs} position. We assume that the \ac{AOA} measurements were all corrupted by clutter in the environment with different standard deviations. Fig. \ref{MIANthing22} (e) compares the \ac{RMSE} performance, this visualization effectively illustrates the impact of angular estimation errors on \ac{RMSE} in the presence of clutter, emphasizing that larger angular offsets lead to significantly higher \ac{RMSE} values. This underscores the importance of precise angle estimation to minimize tracking errors in cluttered environments, which is directly related to the clutter present in the surroundings. When the \ac{APs} have a clear \ac{LoS} to the \ac{UE} without any obstructions, the \ac{AOA} estimation from the radar echo will be more accurate, resulting in a lower \ac{RMSE} value. By ensuring precise angle estimation through strategic AP-UE pairing, the system can minimize tracking errors and enhance the overall performance in challenging cluttered scenarios.
\vspace{-0.3mm}
\par Furthermore, Fig. \ref{MIANthing22} (f) illustrates the impact \ac{AOA} estimation accuracy on the \ac{SCNR} performance. The different curves represent varying degrees of \ac{AOA} estimation error, where 0-degree indicates perfect \ac{AOA} estimation, and 4-degree, 8-degree, and 12-degree denote increasing levels of estimation errors due to clutter in the environment. At lower \ac{SNR} values, the system with perfect \ac{AOA} estimation (0-degree) exhibits the highest \ac{SCNR}, as it can accurately capture the \ac{UEs} radar parameters. However, as the \ac{SNR} increases, the scenarios with larger \ac{AOA} estimation errors start to show degraded \ac{SCNR} performance compared to the perfect estimation case. This highlights the critical importance of precise \ac{AOA} estimation, especially in cluttered environments. When the \ac{APs} have a clear \ac{LoS} to the \ac{UE} without obstructions, the \ac{AOA} estimation from the radar echo will be more accurate, resulting in a higher \ac{SCNR} and lower \ac{RMSE} in target tracking. Conversely, in the presence of clutter, the \ac{AOA} estimation becomes less precise, leading to a lower \ac{SCNR} and higher \ac{RMSE}. 
%%%%%%%%%%%%%%%%%%%%%%%%%%%%%%%%%%%%%%%%%%%%%%%%%%%%%%%%%%%%%%%%%%%%%%%%%%%%%%%%%
\vspace{-3mm}
\subsection{Performance of the Sensing Detector}
\par Figure \ref{MIANTHING} (a) illustrates the performance of the optimal sensing detector in tracking \ac{UE} in the presence of clutter. The velocity-distance plot is based on a binary hypothesis testing framework where the received signal can either consist of clutter and noise $\mathrm{H}_0$ or include the target signal $\mathrm{H}_1$. The optimal detector, derived using the estimator-correlator theorem, is expressed as a canonical form involving the parameters $\lambda_k$ and $\theta_k$. In the figure, the original \ac{UE} location is marked, representing the true position of the target signal. 
As we can see from the figure, due to clutter, the original location of the \ac{UE} was 15m away from the \ac{AP}, but the estimated location shows 45m away, which is a significant error in the range estimation. The estimated \ac{UE} location, offset from the original due to clutter, highlights this error. This graphical representation not only validates the underlying mathematical model but also highlights the detector's sensitivity to clutter, underscoring the necessity for advanced clutter mitigation techniques to enhance tracking accuracy in complex environments.
%%%%%%%%%%%%%%%%%%%%%%%%%%%%%%%%%%%%%%%%%%%%%%%%%%%%%%%%%%%%%%%%%%%%%%%%%%%%%%%%%
\vspace{-3mm}
\subsection{Radar Detection Coverage Probability} 
\par The derived expression for $P_{dc}$ in (\ref{pdcr}) offers critical insights into \ac{UE} sensing/tracking performance in a cluttered environment, validated by Fig. \ref{MIANTHING} (b). This figure shows the impact of clutter on $P_{dc}$ as a function of \ac{SCNR}, highlighting clutter's degrading effect and the importance of the proposed clutter mitigation strategy. The clutter-free curve demonstrates nearly perfect detection at higher \ac{SCNR} levels, with $P_{dc}$ approaching 1 (100\%), as clutter effects ($I(R)$) are absent. Conversely, the curve without the proposed mitigation shows significantly affected detection performance, with lower $P_{dc}$ at lower \ac{SCNR} values, indicating higher false alarm rates and missed detection due to clutter. The reduction in \ac{SCNR} due to clutter is represented by the integral term ($I(R)$) in the $P_{dc}$ expression, accounting for clutter scatterers' spatial distribution and reflectivity. This underscores the necessity of the proposed UE-AP association strategy to enhance detection in cluttered environments.
%%%%%%%%%%%%%%%%%%%%%%%%%%%%%%%%%%%%%%%%%%%%%%%%%%%%%%%%%%%%%%%%%%%%%%%%%%%%%%%%%
\vspace{-3mm}
\subsection{\ac{$P_{dc}$} vs Clutter Density}
\par Figure \ref{MIANTHING} (c) illustrates how the \ac{$P_{dc}$} varies with the distance from the \ac{AP} for different clutter densities $\rho$. The mean clutter cross-section and radar parameters are held constant in this scenario. As clutter density increases, the number of scatterers in the environment also rises, leading to more clutter echoes that interfere with the \ac{AP} sensing detection capabilities. When $\rho=0$, the environment is free from clutter, resulting in the \ac{SCNR} being equivalent to the \ac{SNR}. Consequently, \ac{$P_{dc}$} remains high even at larger distances. However, as 
$\rho$ increases to 0.1, \ac{$P_{dc}$} decreases sharply with increasing distance from the \ac{AP} due to the exponential increase in clutter echoes, which severely impair the \ac{AP} sensing detection performance. From the perspective of our \ac{UA} scheme, this indicates that \ac{UA} needs to account for clutter density to maintain effective sensing performance. As clutter density increases, it becomes crucial to adopt \ac{UA} strategies to ensure that the \ac{UE} can still achieve reliable \ac{$P_{dc}$}. By incorporating clutter parameters into our association decisions, we can enhance sensing accuracy in cluttered environments.
%%%%%%%%%%%%%%%%%%%%%%%%%%%%%%%%%%%%%%%%%%%%%%%%%%%%%%%%%%%%%%%%%%%%%%%%%%%%%%%%%
\vspace{-4mm}
\subsection{\ac{$P_{dc}$} vs Clutter Cross-section}
In Fig. \ref{MIANTHING} (d), the impact of different $\upsilon_{\mathrm{c}_\mathrm{avg}}$ on \ac{$P_{dc}$} is shown for both
single and multiple clutter reflections, at distances of 5 meters and 10 meters. The figure reveals that increasing $\upsilon_\mathrm{c_\mathrm{avg}}$ generally decreases \ac{$P_{dc}$}, as larger clutter cross-sections lead to more significant
radar signal reflections, thus increasing the interference. 
For single clutter reflections, the radar's detection performance at 5 meters is superior to 10 meters due to lower path loss and less accumulated clutter effect. However, in scenarios with multiple clutter reflections, the \ac{$P_{dc}$} deteriorates more rapidly, especially at greater distances. This is due to multiple reflections exacerbating interference, which makes it more challenging for the AP to detect and track the UE. Consequently, the \ac{UA} strategy must prioritize including APs in the cluster that provides higher \ac{$P_{dc}$} while avoiding APs heavily affected by clutter. This ensures more reliable detection and tracking of the UE, optimizing the performance of the \ac{CF-mMIMO} system.
%%%%%%%%%%%%%%%%%%%%%%%%%%%%%%%%%%%%%%%%%%%%%%%%%%%%%%%%%%%%%%%%%%%%%%%%%%%%%%
\vspace{-3mm}
\section{Conclusion} \label{conc}
In conclusion, the proposed \ac{UA} scheme effectively enhances the performance of \ac{JRC} \ac{CF-mMIMO} systems. By seamlessly integrating radar sensing capabilities with the \ac{UA} process, the scheme optimizes the selection of \ac{APs} while mitigating the adverse effects of environmental clutter. This approach addresses the limitations of traditional \ac{UA} methods, which solely consider communication requirements based \ac{AP} selection which in return would not necessarily be optimal for sensing. Simulation results validate the efficiency of the proposed \ac{JRC}-based \ac{UA} scheme, showcasing its ability to significantly outperform traditional approaches. The scheme's scalability and robustness to dynamic environmental conditions ensure that users are associated with \ac{APs} that can provide reliable communication links and accurate target tracking. This innovative approach paves the way for enhanced user experience, increased spectral efficiency, and more reliable target tracking, ultimately contributing to the advancement of integrated radar-communication technologies. The proposed scheme optimizes the user-AP pairings by considering the \ac{RCS}, \ac{SCNR}, and \ac{AOA} estimation accuracy, improving overall \ac{JRC} system performance. Future works may include other parameters of sensing to be incorporated while selecting the \ac{APs} such as sensing resolution, security aspects, and material characterization, etc.
\vspace{-1.9mm}
%%%%%%%%%%%%%%%%%%%%%%%%%%%%%%%%%%%%%%%%%%%%%%%%%%%%%%%%%%%%%%%%%%%%%%%%%%%%%%%%%
{\appendix [Proof of Theorem 2]}
From the definition of the $P_{dc}$, we have:
\begin{equation}
    P_{dc}(r_t) = \mathbb{P}\left[ \mathrm{SCNR}(r_t) \geq \gamma \right].
\end{equation}
Substituting the \ac{SCNR} expression under \ac{NLoS} conditions as
\begin{equation}
P_{dc}(r_t) = \mathbb{P}\left[ \frac{\frac{Z \upsilon_t e^{-2\alpha' r_t}}{r_t^{2q}}}{\mathrm{n}_{l} + \sum_{c \in \Gamma} \frac{Z G(\theta_c) \upsilon_c g_c e^{-2\alpha' r_c}}{r_c^{2q}}} \geq \gamma \right],
\end{equation}
Rearranging the inequality, we obtain
{\small
\begin{equation}
P_{dc}(r_t) = \mathbb{P}\left[ \upsilon_t \geq \frac{\gamma \mathrm{n}_{l} r_t^{2q} e^{2\alpha' r_t}}{Z} + \gamma r_t^{2q} e^{2\alpha' r_t} \sum_{c \in \Gamma} \frac{G(\theta_c) \upsilon_c g_c}{r_c^{2q} e^{2\alpha' r_c}} \right],
\end{equation}
}
using the exponential distribution of $\upsilon_t$ with mean $\upsilon_{\mathrm{t_{avg}}}$ as
{\small
\begin{equation}
\begin{aligned}
&P_{dc}(r_t) = \mathbb{E}_{\upsilon_c, g_c, \Gamma}\\ &\left[ \exp\left( -\frac{\frac{\gamma \mathrm{n}_{l} r_t^{2q} e^{2\alpha' r_t}}{Z} + \gamma r_t^{2q} e^{2\alpha' r_t} \sum_{c \in \Gamma} \frac{G(\theta_c) \upsilon_c g_c}{r_c^{2q} e^{2\alpha' r_c}}}{\upsilon_{\mathrm{t_{avg}}}} \right) \right],
\end{aligned}
\end{equation}
}
separating the exponent, we get
{\small
\begin{equation}
\begin{aligned}
&P_{dc}(r_t) = \exp\left( -\frac{\gamma \mathrm{n}_{l} r_t^{2q} e^{2\alpha' r_t}}{Z \upsilon_{\mathrm{t_{avg}}}} \right) \\ &\mathbb{E}_{\upsilon_c, g_c, \Gamma} \left[ \exp\left( -\frac{\gamma r_t^{2q} e^{2\alpha' r_t}}{\upsilon_{\mathrm{t_{avg}}}} \sum_{c \in \Gamma} \frac{G(\theta_c) \upsilon_c g_c}{r_c^{2q} e^{2\alpha' r_c}} \right) \right],
\end{aligned}
\end{equation}
}
Applying the probability generating functional of the \ac{PPP} $\Gamma$
\begin{equation}
\mathbb{E}\left[ \prod_{c \in \Gamma} f(c) \right] = \exp\left( -\varrho \int_{\mathbb{R}^2} (1 - f(c)) \, dc \right),
\end{equation}
with \( f(c) = \exp\left( -\frac{\gamma r_t^{2q} e^{2\alpha' r_t} G(\theta_c) \upsilon_c g_c}{\upsilon_{\mathrm{t_{avg}}} r_c^{2q} e^{2\alpha' r_c}} \right) \), we get
{\small
\begin{equation}
\begin{aligned}
&P_{dc}(r_t) = \exp\left( -\frac{\gamma \mathrm{n}_{l} r_t^{2q} e^{2\alpha' r_t}}{Z \upsilon_{\mathrm{t_{avg}}}} \right)  \exp\\&\left( -\varrho \int_{\mathbb{R}^2} \left( 1 - \mathbb{E}_{\upsilon_c, g_c} \left[ \exp\left( -\frac{\gamma r_t^{2q} e^{2\alpha' r_t} G(\theta_c) \upsilon_c g_c}{\upsilon_{\mathrm{t_{avg}}} r_c^{2q} e^{2\alpha' r_c}} \right) \right] \right) d\vec{r_c} \right).
\end{aligned}
\end{equation}
}
Assuming $g_c = 1$ (worst-case scenario) and simplifying the expectation inside the integral
{\small
\begin{equation}
\begin{aligned}
&\mathbb{E}_{\upsilon_c} \left[ \exp\left( -\frac{\gamma r_t^{2q} e^{2\alpha' r_t} G(\theta_c) \upsilon_c}{\upsilon_{\mathrm{t_{avg}}} r_c^{2q} e^{2\alpha' r_c}} \right) \right] \\&= \frac{1}{1 + \frac{\gamma r_t^{2q} e^{2\alpha' r_t} G(\theta_c) \upsilon_{\mathrm{c_{avg}}}}{\upsilon_{\mathrm{t_{avg}}} r_c^{2q} e^{2\alpha' r_c}}},
\end{aligned}
\end{equation}
}
we obtain $= \frac{1}{1 + \frac{\nu'(r_t) G(\theta_c)}{r_c^{2q} e^{2\alpha' r_c}}}$, where \( \nu'(r_t) = \frac{\gamma r_t^{2q} e^{2\alpha' r_t} \upsilon_{\mathrm{c_{avg}}}}{\upsilon_{\mathrm{t_{avg}}}} \). Substituting back into the integral, we have
{\small
\begin{equation}
\begin{aligned}
&P_{dc}(r_t) = \exp\left( -\frac{\gamma \mathrm{n}_{l} r_t^{2q} e^{2\alpha' r_t}}{Z \upsilon_{\mathrm{t_{avg}}}} \right)\\& \exp\left( -\varrho \int_0^{2\pi} \int_{r_t}^{r_t + \Delta R} \left( 1 - \frac{1}{1 + \frac{\nu'(r_t) G(\theta_c)}{r_c^{2q} e^{2\alpha' r_c}}} \right) r_c \, dr_c \, d\theta_c \right),
\end{aligned}
\end{equation}
}
Simplifying the inner term
\begin{equation}
1 - \frac{1}{1 + \frac{\nu'(r_t) G(\theta_c)}{r_c^{2q} e^{2\alpha' r_c}}} = \frac{\nu'(r_t) G(\theta_c)}{\nu'(r_t) G(\theta_c) + r_c^{2q} e^{2\alpha' r_c}},
\end{equation}
we get
{\small
\begin{equation}
\begin{aligned}
&P_{dc}(r_t) = \exp\left( -\frac{\gamma \mathrm{n}_{l} r_t^{2q} e^{2\alpha' r_t}}{Z \upsilon_{\mathrm{t_{avg}}}} \right)\\& \exp\left( -\varrho \int_0^{2\pi} \int_{r_t}^{r_t + \Delta R} \frac{\nu'(r_t) G(\theta_c) r_c}{\nu'(r_t) G(\theta_c) + r_c^{2q} e^{2\alpha' r_c}} \, dr_c \, d\theta_c \right)
\end{aligned}
\end{equation}
}
Combining the exponents, we have the final result
\begin{equation}
P_{dc}(r_t) = J(r_t) \exp\left( -\frac{\gamma \mathrm{n}_{l} r_t^{2q} e^{2\alpha' r_t}}{Z \upsilon_{\mathrm{t_{avg}}}} \right),
\end{equation}
\vspace{-1mm}
where
{\small
\begin{equation}
\begin{aligned}
&J(r_t) = \\&\exp\left( -\varrho \int_0^{2\pi} \int_{r_t}^{r_t + \Delta R} \frac{\nu'(r_t) G(\theta_c) r_c}{\nu'(r_t) G(\theta_c) + r_c^{2q} e^{2\alpha' r_c}} \, dr_c \, d\theta_c \right)
\end{aligned}
\end{equation}
}

%%%%%%%%%%%%%%%%%%%%%%%%%%%%%%%%%%%%%%%%%%%%%%%%%%%%%%%%%%%%%%%%%%%%%%%%%%%%%%%%%
% Final Comments:
%    Abstract DONE
%    Introduction DONE (Just check Notations)
%    System Model DONE
%    Problem Formulation
%    Unlike \cite{behdad2024multi}, where each \ac{AP} is either a \ac{JRC} transmitter or a sensing receiver, our approach enables each \ac{AP} to function as both.

%%%%%%%%%%%%%%%%%%%%%%%%%%%%%%%%%%%%%%%%%%%%%%%%%%%%%%%%%%%%%%%%%%%%%%%%%%%%%%%%%



\begin{thebibliography}{10}
\providecommand{\url}[1]{#1}
\csname url@samestyle\endcsname
\providecommand{\newblock}{\relax}
\providecommand{\bibinfo}[2]{#2}
\providecommand{\BIBentrySTDinterwordspacing}{\spaceskip=0pt\relax}
\providecommand{\BIBentryALTinterwordstretchfactor}{4}
\providecommand{\BIBentryALTinterwordspacing}{\spaceskip=\fontdimen2\font plus
\BIBentryALTinterwordstretchfactor\fontdimen3\font minus \fontdimen4\font\relax}
\providecommand{\BIBforeignlanguage}[2]{{%
\expandafter\ifx\csname l@#1\endcsname\relax
\typeout{** WARNING: IEEEtran.bst: No hyphenation pattern has been}%
\typeout{** loaded for the language `#1'. Using the pattern for}%
\typeout{** the default language instead.}%
\else
\language=\csname l@#1\endcsname
\fi
#2}}
\providecommand{\BIBdecl}{\relax}
\BIBdecl

\bibitem{wei2023integrated}
Z.~Wei \emph{et~al.}, ``Integrated sensing and communication signals toward 5{G}-{A} and 6{G}: A survey,'' \emph{IEEE Internet Things J.}, vol.~10, no.~13, pp. 11\,068--11\,092, 2023.

\bibitem{andrews2016we}
J.~G. Andrews \emph{et~al.}, ``Are we approaching the fundamental limits of wireless network densification?'' \emph{IEEE Communications Magazine}, vol.~54, no.~10, pp. 184--190, 2016.

\bibitem{8258595}
S.~Parkvall \emph{et~al.}, ``{NR}: The new 5{G} radio access technology,'' \emph{IEEE Communications Standards Magazine}, vol.~1, no.~4, pp. 24--30, 2017.

\bibitem{chen2022survey}
S.~Chen \emph{et~al.}, ``A survey on user-centric cell-free massive {MIMO} systems,'' \emph{Digital Communications and Networks}, vol.~8, no.~5, pp. 695--719, 2022.

\bibitem{demir2021foundations}
{\"O}.~T. Demir \emph{et~al.}, ``Foundations of user-centric cell-free massive {MIMO},'' \emph{Foundations and Trends{\textregistered} in Signal Processing}, vol.~14, no. 3-4, pp. 162--472, 2021.

\bibitem{ngo2015cell}
H.~Q. Ngo \emph{et~al.}, ``Cell-free massive {MIMO}: Uniformly great service for everyone,'' in \emph{IEEE 16th international workshop on signal processing advances in wireless communications}.\hskip 1em plus 0.5em minus 0.4em\relax IEEE, 2015, pp. 201--205.

\bibitem{10411070}
J.~Fang \emph{et~al.}, ``Cell-free {mMIMO} systems in short packet transmission regime: Pilot and power allocation,'' \emph{IEEE Trans. Veh. Technol.}, vol.~73, no.~6, pp. 8322--8337, 2024.

\bibitem{bjornson2020scalable}
E.~Bj{\"o}rnson \emph{et~al.}, ``Scalable cell-free massive {MIMO} systems,'' \emph{IEEE Trans. Commun.}, vol.~68, no.~7, pp. 4247--4261, 2020.

\bibitem{8000355}
S.~Buzzi and C.~D’Andrea, ``Cell-free massive {MIMO}: User-centric approach,'' \emph{IEEE Wireless Commun. Lett.}, vol.~6, no.~6, pp. 706--709, 2017.

\bibitem{chen2020structured}
S.~Chen \emph{et~al.}, ``Structured massive access for scalable cell-free massive {MIMO} systems,'' \emph{IEEE J. Sel. Areas Commun.}, vol.~39, no.~4, pp. 1086--1100, 2020.

\bibitem{ngo2018performance}
H.~Q. Ngo \emph{et~al.}, ``On the performance of cell-free massive {MIMO} in ricean fading,'' in \emph{52nd Asilomar Conference on Signals, Systems, and Computers}.\hskip 1em plus 0.5em minus 0.4em\relax IEEE, 2018, pp. 980--984.

\bibitem{sarker2023access}
M.~Sarker \emph{et~al.}, ``Access point-user association and auction algorithm-based pilot assignment schemes for cell-free massive {MIMO} systems,'' \emph{IEEE Syst. J.}, vol.~17, no.~3, pp. 4301--4312, 2023.

\bibitem{d2020user}
C.~D’Andrea \emph{et~al.}, ``User association in scalable cell-free massive {MIMO} systems,'' in \emph{54th Asilomar Conference on Signals, Systems, and Computers}.\hskip 1em plus 0.5em minus 0.4em\relax IEEE, 2020, pp. 826--830.

\bibitem{behdad2024multi}
Z.~Behdad \emph{et~al.}, ``Multi-static target detection and power allocation for integrated sensing and communication in cell-free massive {MIMO},'' \emph{IEEE Trans. Wireless Commun.}, 2024.

\bibitem{behdad2024joint}
Z.~Behdad, {\"O}.~T. Demir \emph{et~al.}, ``Joint processing and transmission energy optimization for {ISAC} in cell-free massive {MIMO} with {URLLC},'' \emph{arXiv preprint arXiv:2401.10315}, 2024.

\bibitem{elfiatoure2024multiple}
M.~Elfiatoure \emph{et~al.}, ``Multiple-target detection in cell-free massive {MIMO}-assisted {ISAC},'' \emph{arXiv preprint arXiv:2404.17263}, 2024.

\bibitem{mao2023beamforming}
W.~Mao \emph{et~al.}, ``Beamforming design in cell-free massive {MIMO} integrated sensing and communication systems,'' in \emph{GLOBECOM IEEE Global Communications Conference}.\hskip 1em plus 0.5em minus 0.4em\relax IEEE, 2023, pp. 546--551.

\bibitem{mao2024communication}
W.~Mao, Lu \emph{et~al.}, ``Communication-sensing region for cell-free massive {MIMO} {ISAC} systems,'' \emph{IEEE Trans. Wireless Commun.}, 2024.

\bibitem{9940605}
J.~Xu \emph{et~al.}, ``Hybrid index modulation for dual-functional radar communications systems,'' \emph{IEEE Trans. Veh. Technol.}, vol.~72, no.~3, pp. 3186--3200, 2023.

\bibitem{9724187}
Z.~Xiao \emph{et~al.}, ``Waveform design and performance analysis for full-duplex integrated sensing and communication,'' \emph{IEEE J. Sel. Areas Commun.}, vol.~40, no.~6, pp. 1823--1837, 2022.

\bibitem{10058895}
X.~Ma \emph{et~al.}, ``Cooperative beamforming for {RIS}-aided cell-free massive {MIMO} networks,'' \emph{IEEE Trans. Wireless Commun.}, vol.~22, no.~11, pp. 7243--7258, 2023.

\bibitem{9713691}
S.~Gopi \emph{et~al.}, ``Cooperative {3D} beamforming for small-cell and cell-free {6G} systems,'' \emph{IEEE Trans. Veh. Technol.}, vol.~71, no.~5, pp. 5023--5036, 2022.

\bibitem{bjornson2017massive}
E.~Bj{\"o}rnson \emph{et~al.}, ``Massive {MIMO} networks: Spectral, energy, and hardware efficiency,'' \emph{Foundations and Trends{\textregistered} in Signal Processing}, vol.~11, no. 3-4, pp. 154--655, 2017.

\bibitem{ngo2017cell}
H.~Q. Ngo \emph{et~al.}, ``Cell-free massive {MIMO} versus small cells,'' \emph{IEEE Trans. Wireless Commun.}, vol.~16, no.~3, pp. 1834--1850, 2017.

\bibitem{article}
E.~Björnson \emph{et~al.}, ``Massive {MIMO} networks: Spectral, energy, and hardware efficiency,'' \emph{Foundations and Trends® in Signal Processing}, vol.~11, pp. 154--655, 01 2017.

\bibitem{10216314}
J.~Park \emph{et~al.}, ``Downlink cell-free massive {MIMO} with pilot contamination,'' \emph{IEEE Trans. Veh. Technol.}, vol.~73, no.~1, pp. 1412--1417, 2024.

\bibitem{10415170}
A.~Sesyuk \emph{et~al.}, ``Radar-based millimeter-wave sensing for accurate 3-{D} indoor positioning: Potentials and challenges,'' \emph{IEEE Journal of Indoor and Seamless Positioning and Navigation}, vol.~2, pp. 61--75, 2024.

\bibitem{7869024}
E.~Nayebi \emph{et~al.}, ``Performance of cell-free massive {MIMO} systems with {MMSE} and {LSFD} receivers,'' in \emph{2016 50th Asilomar Conference on Signals, Systems and Computers}, 2016, pp. 203--207.

\bibitem{chiu2013stochastic}
S.~N. Chiu \emph{et~al.}, \emph{Stochastic geometry and its applications}.\hskip 1em plus 0.5em minus 0.4em\relax John Wiley \& Sons, 2013.

\bibitem{skolnik1980introduction}
M.~I. Skolnik \emph{et~al.}, \emph{Introduction to radar systems}.\hskip 1em plus 0.5em minus 0.4em\relax McGraw-hill New York, 1980, vol.~3.

\bibitem{nguyen2013new}
T.~L.~N. Nguyen \emph{et~al.}, ``A new approach for positioning based on {AOA} measurements,'' in \emph{International Conference on Computing, Management and Telecommunications}.\hskip 1em plus 0.5em minus 0.4em\relax IEEE, 2013, pp. 208--211.

\bibitem{7346436}
J.~Yin \emph{et~al.}, ``A simple and accurate {TDOA-AOA} localization method using two stations,'' \emph{IEEE Signal Process. Lett.}, vol.~23, no.~1, pp. 144--148, 2016.

\bibitem{10458884}
A.~Naeem \emph{et~al.}, ``Polarization-based multiplexing: Enabling spectrum efficient joint radar and communication,'' \emph{IEEE Wireless Commun. Lett.}, vol.~13, no.~5, pp. 1414--1418, 2024.

\bibitem{fish1909coordinates}
J.~C.~L. Fish, \emph{Coordinates of elementary surveying}.\hskip 1em plus 0.5em minus 0.4em\relax The author, 1909.

\bibitem{stewart2011calculus}
J.~Stewart, \emph{Calculus, Early Transcendentals, AP ed.}\hskip 1em plus 0.5em minus 0.4em\relax Cengage Learning, 2011.

\bibitem{10310167}
X.~Qiao \emph{et~al.}, ``Two-layer large scale fading precoding for cell-free massive {MIMO}: Performance analysis and optimization,'' \emph{IEEE Trans. Veh. Technol.}, vol.~73, no.~3, pp. 3901--3916, 2024.

\bibitem{6601713}
M.~M. Naghsh \emph{et~al.}, ``Unified optimization framework for multi-static radar code design using information-theoretic criteria,'' \emph{IEEE Trans. Signal Process.}, vol.~61, no.~21, pp. 5401--5416, 2013.

\bibitem{kay1998statistical}
S.~M. Kay, ``Statistical signal processing, volume {II}, detection theory,'' 1998.

\bibitem{3GPP2017}
{3GPP}, ``Further advancements for {E-UTRA} physical layer aspects (release 9),'' 3GPP, Tech. Rep. TS 36.814, 2017.

\bibitem{guruacharya2020map}
S.~Guruacharya \emph{et~al.}, ``{MAP} ratio test detector for radar system,'' \emph{IEEE Trans. Signal Process.}, vol.~69, pp. 573--588, 2020.

\bibitem{9174860}
S.~Chen \emph{et~al.}, ``Structured massive access for scalable cell-free massive {MIMO} systems,'' \emph{IEEE J. Sel. Areas Commun.}, vol.~39, no.~4, pp. 1086--1100, 2021.

\bibitem{9449014}
M.~Sarker \emph{et~al.}, ``Granting massive access by adaptive pilot assignment scheme for scalable cell-free massive {MIMO} systems,'' in \emph{IEEE 93rd Vehicular Technology Conference}, 2021, pp. 1--5.

\end{thebibliography}
\end{document}